\documentclass[12pt,preprint]{aastex}
\begin{document}
\title{Evidence of the Dynamics of Relativistic Jet Launching in Quasars}
\author{Brian Punsly\altaffilmark{1}}
\altaffiltext{1}{1415 Granvia Altamira, Palos Verdes Estates CA, USA
90274 and ICRANet, Piazza della Repubblica 10 Pescara 65100, Italy,
brian.punsly1@verizon.net}
\begin{abstract}
Hubble Space Telescope (HST) spectra of the extreme ultraviolet
(EUV), the optically thick emission from the innermost accretion
flow onto the central supermassive black hole, indicate that RLQs
tend to be EUV weak compared to the radio quiet quasars (RQQs); yet
the remainder of the optically thick thermal continuum is
indistinguishable. The deficit of EUV emission in RLQs has a
straightforward interpretation as a missing or a suppressed
innermost region of local energy dissipation in the accretion flow.
This article is an examination of the evidence for a distribution of
magnetic flux tubes in the innermost accretion flow that results in
magnetically arrested accretion (MAA) and creates the EUV deficit.
These same flux tubes and possibly the interior magnetic flux that
they encircle are the source of the jet power as well. In the MAA
scenario, islands of large scale magnetic vertical flux perforate
the innermost accretion flow of RLQs. The first prediction of the
theory that is supported by the HST data is that the strength of the
(large scale poloidal magnetic fields) jets in the MAA region is
regulated by the ram pressure of the accretion flow in the quasar
environment. The second prediction that is supported by the HST data
is that the rotating magnetic islands remove energy from the
accretion flow as a Poynting flux dominated jet in proportion to the
square of the fraction of the EUV emitting gas that is displaced by
these islands.
\end{abstract}


\section{Introduction}
The mechanism that drives powerful beams of radio emitting plasma,
moving near the speed of light, in radio loud quasars (RLQs) has
been the subject of much speculation in the literature, with little
or nothing in the way of observations of the region that launches
the jets \citep{lov76,bla77,bla82,pun08}. This circumstance has now
changed based on Hubble Space Telescope (HST) observations of the
extreme ultraviolet (EUV) in quasars \citep{pun14}. Quasars are
generally associated with the optically thick thermal emission from
gas that accretes onto a supermassive black hole
\cite{lyn71,sha73,nov73,mal83,szu96}. Curiously, $\sim 10\%$ of
these quasars have conspicuous beams, or jets, of relativistic
plasma on scales that can exceed a million light years and powers $
10^{3}$ - $10^{4}$ the integrated light of the largest galaxies,
RLQs. The optical/ultraviolet spectra of RLQs and radio quiet
quasars (RQQs) tend to be very similar except for subtle differences
in certain emission line strengths and widths
\citep{ste91,bor92,bro94}. These emission line regions are far from
the central engine, $\sim$ $10^{3}$ - $10^{4}$ larger than the
central black hole radius. Consequently, this research path has
provided very little understanding of the jet launching mechanism.
The EUV continuum, $\lambda < 1100$ \AA\,, is created orders of
magnitude closer to the central engine and RLQs display a
significant EUV continuum deficit relative to RQQs \citep{tel02}.

\par The quasar luminosity is widely believed to arise from the viscous dissipation of
turbulence driven by the differential rotational shearing of
accreting gas \citep{sha73}. In numerical and theoretical models,
the highest frequency optically thick thermal emission arises from
the innermost region of the accretion flow and its frequency is
shortward of the peak of the spectral energy distribution (SED)
\citep{zhu12}. Consider this in the context of the RLQ and RQQ
quasar composite spectra from HST data in Figure 1 \citep{tel02}.
The continuum of the composite spectra are indistinguishable except
for the EUV emission shortward of the peak of the SED at
$\approx\lambda =1100 \AA$ (where the spectra are normalized to 0).
Thus, the difference in the EUV emission between RLQs and RQQs
likely arises from suppressed emission in the innermost region of
the accretion flow in RLQs of what is otherwise a similar accretion
flow to that found in RQQs \citep{pun14}.

\par In this article, the explicit predictions of a MAA (magnetically arrested accretion)
description of the EUV deficit in RLQs are explored (see Figure 2
for a schematic illustration). It is posited that islands of large
scale magnetic vertical flux perforate the accretion flow of RLQs
within a few black hole radii of the central black hole
\cite{pun14,igu03,igu08}. Three pieces of information are
synthesized to constrain the dynamics in the inner accretion flow,
the long term time average jet power, $Q$, the peak luminosity of
the SED (a good surrogate for the bolometric luminosity,
$L_{\mathrm{bol}}$, of the accretion flow) and the EUV deficit. MAA
indicates that the rotating magnetic islands remove energy from the
accretion flow as a Poynting flux dominated jet in proportion to the
square of the fraction of the EUV emitting gas that is displaced by
these islands and in direct proportion to the accretion rate. These
predicted relationships are well fit by the data extracted from the
HST spectra, thus providing strong support for the MAA
interpretation of RLQs and the paradigm that magnetic flux in the
inner accretion flow is the switch that launches quasar jets
\citep{mei99}.
\begin{figure*}
\includegraphics[width=170 mm, angle= 0]{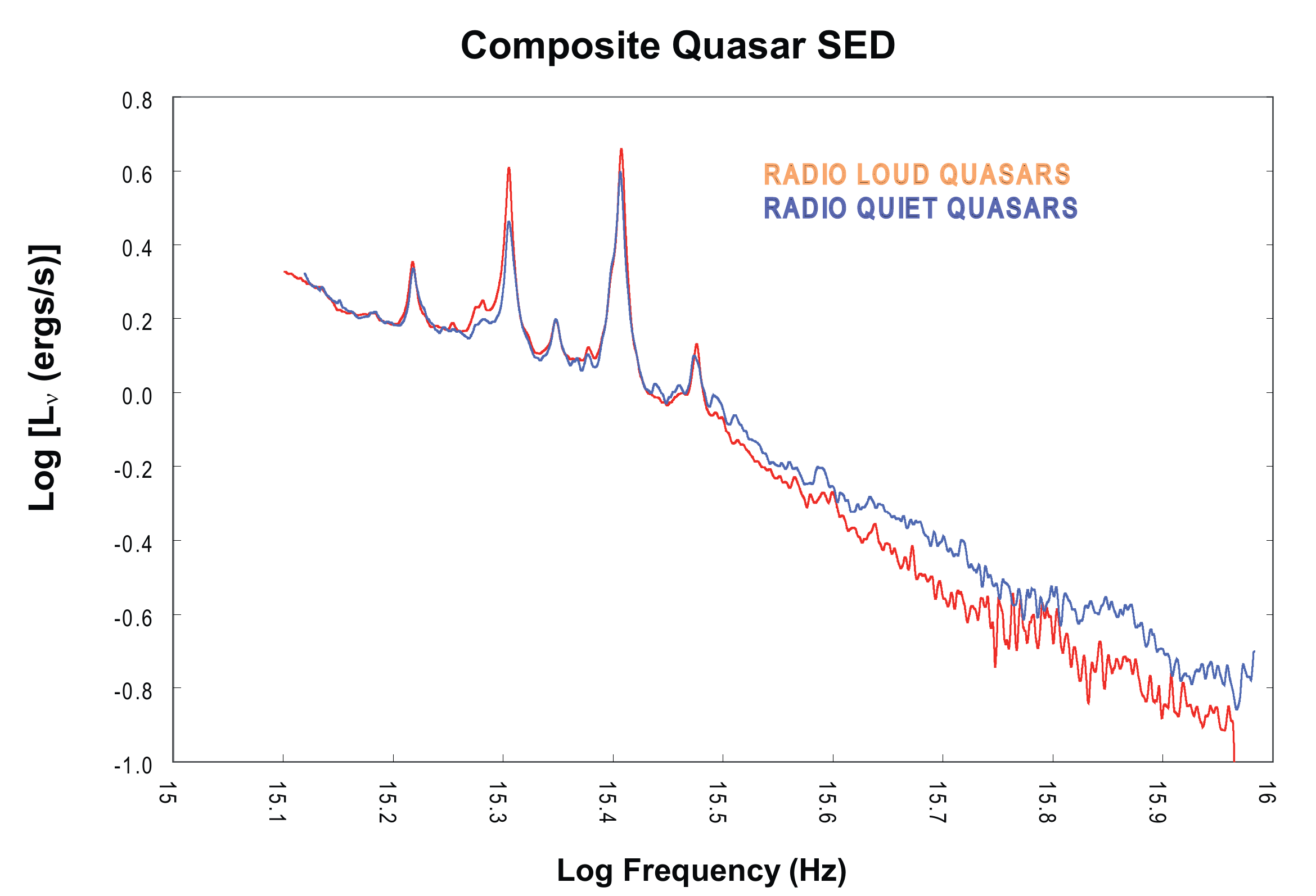}
\caption{The EUV Deficit. The HST quasar composite continuum
accretion disk spectra. The blue and red plots are the ultraviolet
SED peak and the EUV composite spectra for RQQs and RLQs,
respectively (Telfer et al 2002). Note the steeper SED (larger
$\alpha_{\mathrm{EUV}}$) for RLQs.}
\end{figure*}
\par In the following an extended magnetically arrested scenario is
explored (EMAA) in parallel with the analysis of the MAA scenario.
In the EMAA model, the magnetic flux that threads the EUV region is
the outermost extent of an interior magnetic flux distribution that
threads the equatorial plane of the ergosphere and the event horizon
of the central supermassive black hole. The one necessary assumption
that is required to connect this interior region to observation is
that the vertical magnetic flux in the interior region is
proportional to the magnetic flux in the EUV region, $\Phi_{int}
\propto \Phi_{EUV}$. In this circumstance, the interior magnetic
flux distribution is a major source of jet power. The EUV deficit is
created by the flux that perforates the innermost accretion flow.
Based on the assumption above, this interaction also provides an
indirect tracer (a probe with a linear response function) of
$\Phi_{int}$. With this assumption, the observational data discussed
here also supports the EMAA model.
\par The paper is structured as follows. The second section is a
review of MAA and EUV suppression. In this section, the fundamental
relationships between jet power and EUV suppression are found. The
goal of this article is to test these predicted relationships versus
observations. In order to define the experiment, a method for
estimating the jet power from the radio data and a sample of quasars
with suitable HST spectra and radio images need to be established.
These are the subjects of Sections 3 and 4. Section 5 describes the
experimental tests performed and the conclusions. The last section
is a discussion of the results.
\begin{figure*}
\includegraphics[width=170 mm, angle= 0]{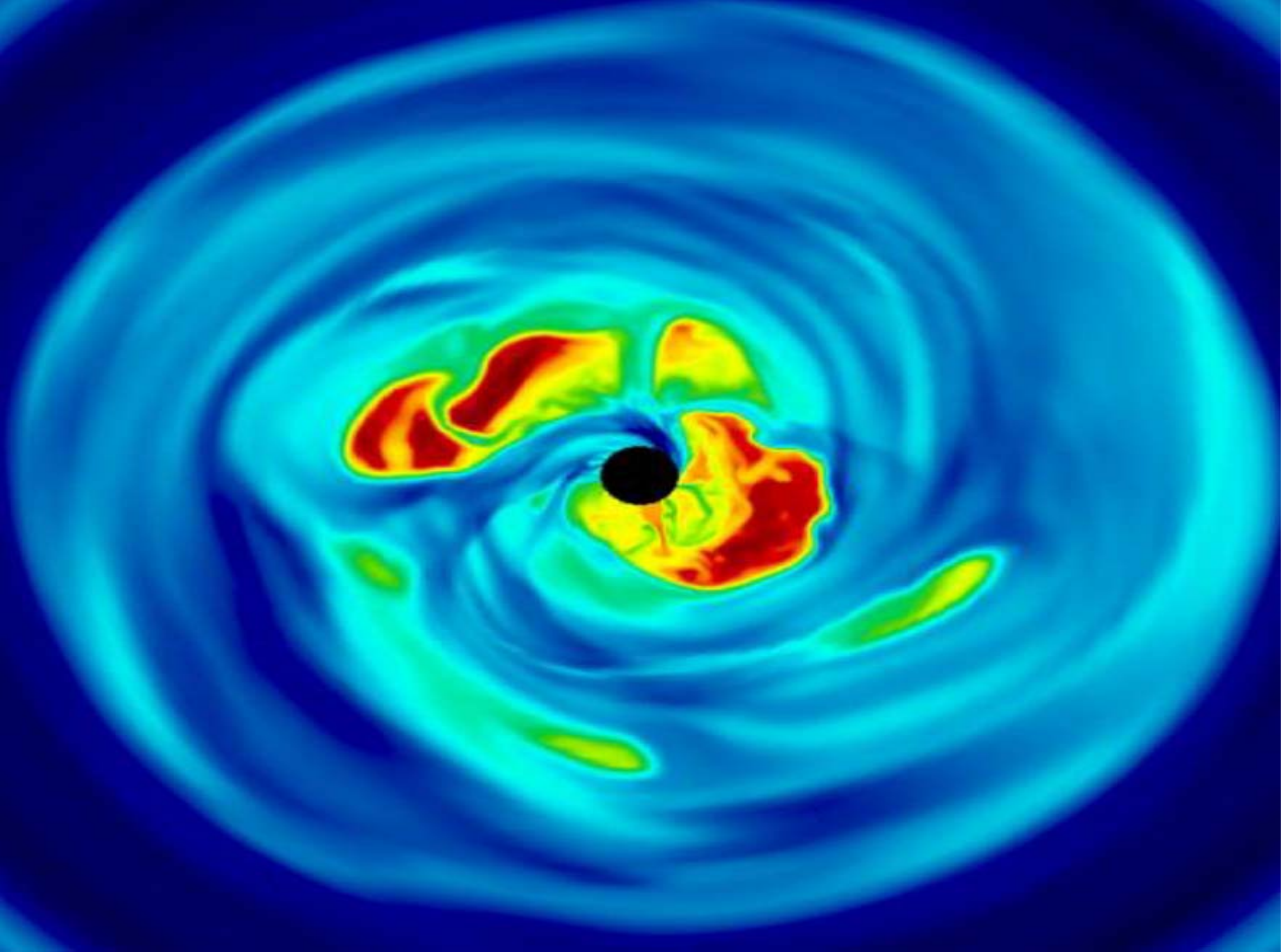}
\caption{Magnetically Arrested Accretion. The logarithmic false
color contour plot of the vertical poloidal magnetic pressure (i.e.,
$B_{p} = B_{z}$) from an MAA simulation is used to schematically
illustrate the concepts posited in this study. One can see Figure 2
of Punsly et al (2009) for the color scale, but this is not
necessary for this discussion. The false color accentuates the
two-fluid nature of the accretion flow. The red to yellow regions
are magnetic islands, the origin of strong Poynting jets and weak
EUV emission. The dark blue (the densest plasma) and green are
strong EUV emitting regions.}
\end{figure*}

\section{Jet Production and Suppressed EUV} The dynamics of viscous shear
in black hole accretion flows with and without MAA are reviewed in
order to assess the affects of MAA in an annular ring of the
accretion flow surrounding the central black hole. Assume that
magnetic islands fill a fraction, $f_{V}$, of the volume of the
ring, $V$, and penetrate a fraction, $f$, of the top and bottom
surface areas of the annular volume, $SA$.  The volume of magnetic
islands is $V_{MI}$ and its complement in $V$ is $V_{MI}^{C}$,
$f_{V} = \int{dV_{MI}}/V$. The surface area elements of the top and
bottom faces are, $dS\!A_{MI}$ and $dS\!A_{MI}^{C}$, respectively.
The turbulent dissipation that heats the plasma in accretion flows
is produced as a consequence of the magneto-rotational instability
(MRI) in the 3-D numerical simulations \cite{dev03,pen10}. The
accretion flow in MAA simulations is perforated by large scale
magnetic flux tubes, magnetic islands (see Figure 2). Thus, MAA
creates a two component fluid. Firstly, there are regions in which
energy is removed from the fluid by turbulent dissipation. Secondly,
in the magnetic islands, energy is removed from the flow vertically
as Poynting flux \citep{igu03,igu08,mck12,tch11}. The large scale
magnetic flux suppresses the MRI induced dissipation in these
regions. Simulated MAA flows are subsonic and therefore do not
produce significant gas heating from shocks \citep{mck12,pun14}.
Thus, viscous dissipation is the primary source of heat creation at
the boundary of the magnetic islands and in surrounding accreting
gas. The total volume available for MRI induced viscous heating is
reduced by the magnetic islands. MAA provides an alternative to
local turbulent dissipation for removing energy from the plasma
allowing the gas to accrete closer to the black hole. The magnetic
islands radiate Poynting flux, $S^{P}$, along the magnetic field
lines at the expense of the energy of the plasma \citep{igu08}. The
local physics that produces the turbulent viscosity, $\eta_{t}$, in
$V_{MI}^{C}$ is unchanged from standard accretion and therefore so
is the viscous stress in the fluid surrounding the islands, $T_{r
\phi}^{\mathrm{visc}}= \eta_{t}r(d\Omega/dr)$. It was shown in
\citet{pun14}, in the magnetically arrested case, the radiative
luminosity is $\approx 1-f$ of what it would be for standard
accretion with the same mass accretion rate. If $f$ is the fraction
of the inner accretion flow surface area, $SA$, penetrated by
magnetic islands in MAA then the EUV luminosity, $L(\mathrm{EUV})$,
obeys the approximate scaling \citet{pun14}
\begin{equation}
 L(\mathrm{EUV}) \propto (1-f)SA \;.
\end{equation}

\par For any MHD Poynting flux dominated jet, regardless of the
source, the total integrated electromagnetic poloidal energy flux is
\begin{equation}
\int{S^{P}\mathrm{d}A_{_{\perp}}} =
k\frac{\Omega_{F}^{2}\Phi^{2}}{2\pi^{2} c}\;,
\end{equation}
where $\Phi$ is the total magnetic flux enclosed within the jet,
$\mathrm{d}A_{_{\perp}}$ is the cross-sectional area element and $k$
is a geometrical factor that equals 1 for a uniform highly
collimated jet \citep{pun08}. Thus, not only do the magnetic islands
of large scale poloidal flux in the inner accretion flow suppress
radiation from this region, but they provide a source of Poynting
flux (power for the jet) as they orbit around the black hole with an
angular velocity, $\Omega_{\mathrm{F}}$. If $M$ is the black hole
mass in geometrized units, $\Omega_{F}\sim c/M$ for rotation based
on black hole spin or Keplerian orbits. Thus, Equation (2) implies
for magnetic islands
\begin{equation}
Q\approx \int S^{P}  \mathrm{d}A_{_{\perp}} \propto
B_{z}^{2}f^{2}\,SA \propto f^{2}P_{\mathrm{ram}}\,SA\ \propto
f^{2}\dot{M}(v_{r}/\theta) \;,
\end{equation}
where $B_{z}$ is the vertical magnetic field at the disk surface.
The string of proportionality statements in Equation (3) require
elaboration. The first proportionality arises from the fact that
magnetic flux, $\Phi$ is conserved in the perfect
magnetohydrodynamic approximation. So the value in the jet is the
same as that in the MAA region. In the MAA region $\Phi
=\int{B_{z}dS\!A_{MI}}$. This integral is approximated as $\Phi
\approx fB_{z}SA$. At this point it is worth reiterating the intent
of exploring the EMAA model as well. Equations (2) and (3) can be
evaluated using $\Phi\approx \Phi_{int}$ for the jet power, and
$\Phi_{int}\propto \Phi_{EUV} \approx fB_{z}SA$. Thus, the following
analysis can be used to compare the pure MAA scenario or the EMAA
scenario with observation. The only qualification is that the EMAA
analysis has the additional assumption that $\Phi_{int}$ scales
linearly on average with $\Phi_{EUV}$.
\par The second proportionality derives from pressure
balance at the interface of the magnetic islands and the enveloping
accretion flow. In order to evaluate the pressure balance, it is
important to realize that the magnetic islands are not frozen into
the enveloping accretion flow. For example, in Figure 2 the magnetic
islands are in the process of spiraling outward, slowly. The
dynamics are not time stationary. Outward motion of the stronger
magnetic islands begins in the innermost regions of the accretion
flow where the magnetic islands tend to merge \citep{igu08}.
Furthermore, the gas density in the magnetic islands decreases due
to outflow (i.e., a jet) in the inner accretion flow. The density is
$<1\%$ of the enveloping accretion flow \citep{igu08}. Hence, they
are weakly affected by gravity. By contrast, the dense enveloping
accretion flow is strongly attracted to the central black hole. This
is a classic Kruskal- Schwarzschild instability, the magnetized
version of a Rayleigh-Taylor instability \citep{sti92}. Thus, the
magnetic islands become buoyant and drift outward relative to the
radial ram pressure of the inflow as discussed in \citet{pun09} and
the associated video of a simulated flow. The magnetic field in the
islands decreases (expansion and ablation) as they drift outward in
response to the decreasing dynamic pressure in  an attempt to
achieve a new pressure balance with the enveloping medium
\citep{igu08}. This dynamic is unsuccessful and the islands slowly
drift outward. In the MAA model considered here, the magnitude of
the outward migration velocity, $v_{MI}^{r}$ is much smaller than
the magnitude of the bulk radial velocity of the inward enveloping
accretion flow, $v_{r}$. This is generally true in the simulations
of \citet{igu08}, except at the smallest radii where the stronger
islands are sometimes particularly unstable and burst outward until
a more stable equilibrium is achieved. By contrast, the vertical
magnetic flux that permeates the inner accretion flow in the
simulation of the Kerr geometry discussed in detail in
\citet{pun09,haw06} seems to persist at the smallest radii for
longer than it does in the simulations of \citet{igu08}, remaining
within $r < 2M$ (in geometrized units), the ergosphere, for $\sim 2$
local Keplerain orbital periods \citet{pun07}. Designate quantities
in the magnetic island by the label MI and in the enveloping
accretion flow by AF. There are four pressure components, $P_{g}$,
$P_{B}$, $P_{r}$, and $P_{\mathrm{ram}}$, corresponding to gas,
magnetic, radiation and ram pressure, respectively. $P_{r}$ is a
slowly varying quantity around horizontal closed loops in a local
neighborhood of the optically thick accretion flow that encircles
each magnetic island. The radiation pressure is continuous through
the low density magnetic islands. Thus, from the second moment of
the radiative transfer equation, one expects $P_{r}(MI) \approx
P_{r}(AF)$ at the interface \citep{pun96}. In the inner accretion
flow, $P_{\mathrm{ram}} \gg P_{g}$ in luminous quasars. Unlike
$P_{r}$, the ram pressure is not continuous across the boundary,
$P_{\mathrm{ram}}(MI) \ll P_{\mathrm{ram}}(AF)$, because of the
large density differential and the motion of the magnetic islands
relative to the enveloping accretion flow \citep{pun96}. There is
also no buoyancy force in the azimuthal direction. The force on the
magnetic islands imposed by the inflow of matter is the strongest
force resisting the buoyant outflow of the magnetic islands. Thus,
despite all the uncertainty in the precise physics of magnetic
island time evolution (see the final section), the radial ram
pressure should determine the internal magnetic pressure of the
islands to first approximation. Therefore, the pressure balance at
the interface of the magnetic island and the enveloping accretion
flow is approximately $P_{\mathrm{ram}}(AF) \approx P_{B}(MI)$
\citep{igu08}. Consequently, the substitution $B_{z}^{2} \propto
P_{\mathrm{ram}}$ in Equation (2) was made.  In summary, if the
magnetic islands are not extremely short-lived transient features in
the inner accretion flow, an approximate balance of the ram pressure
and the magnetic pressure of the poloidal magnetic field in the
islands must exist.
\par The third proportionality on the right hand side of Equation (2) arises from
mass conservation near the black hole, where the ``pseudo-half
angle", $\theta$, is the ratio of cross sectional area to $SA$
(evaluated near the outer boundary of the MAA region). Since, as
discussed above,  $v_{r}- v_{MI}^{r} \approx v_{r}$, to the accuracy
of scaling laws in Equation (3), $P_{\mathrm{ram}} \propto
v_{r}^{2}$. Consider this in the context of the bolometric
luminosity of the accretion disk, $L_{\mathrm{bol}}
=\eta(a)\dot{M}c^{2}$, where $\dot{M}c^{2}$ is the mass-energy
accretion rate and the accretion efficiency is $\eta(a)$, a function
of black hole spin, $a$. Equation (3) can be transformed into an
approximate relationship that is more conducive to comparison with
observation
\begin{equation}
Q/L_{\mathrm{bol}} \approx C(1/\eta(a))(v_{r}/\theta) f^{2} \;,
\end{equation}
where C is a constant. For simplicity, it is assumed that the vast
majority of black holes in quasars are rapidly rotating
\citep{bar70,elv02}. Secondly, it is also assumed that variations in
$\theta$ and $v_{r}$ provide random scatter to Equation (4) and are
not the primary physical drivers of $Q$. It is implicit in the
approximate form of Equation (4) that the exact distributions of
magnetic islands and the associated functional variation of
$\Omega_{F}$ in the magnetically arrested region near the black hole
are higher order corrections and only contribute cosmic scatter to
the simplified relationship. These approximations are collectively
referred to as the homogeneous approximation for the MAA. This makes
the following analysis tractable and not dependent of specific
models of the various parameters. With these assumptions, Equation
(4) and the scaling with $f$ of $L(\mathrm{EUV})$ in Equation (1)
implies an approximate relationship,
\begin{equation}
Q/L_{\mathrm{bol}} \approx A(1 -
L(\mathrm{EUV})/L(\mathrm{EUV})_{\mathrm{RQQ}})^{2} \;,
\end{equation}
where A is a constant and $L(\mathrm{EUV})_{\mathrm{RQQ}}$ is the
fiducial EUV luminosity if there were no magnetic islands.
\section{Long Term Time Averaged Jet Power.} The two most viable options for estimating the
jet power, $Q$, of quasars are either based on the low frequency
(151 MHz) flux from the radio lobes on 100 kpc scales or models of
the broadband Doppler boosted synchrotron and inverse Compton
radiation spectra associated with the relativistic parsec scale jet.
Each method has its advantage. The 151 MHz method is generally
considered more reliable since it does not involve large
uncertainties due to Doppler beaming \citep{wil99}. A disadvantage
is that it involves long term time averages, $\overline{Q}$, that do
not necessarily reflect the current state of quasar activity. In
this study $Q \equiv \overline{Q}$. A method that allows one to
convert 151 MHz flux densities, $F_{151}$ (measured in Jy), into
estimates of long term time averaged jet power, $\overline{Q}$,
(measured in ergs/s) is captured by the formula derived in
\citet{wil99,pun05}:
\begin{eqnarray}
 && \overline{Q} \approx [(\mathrm{\textbf{f}}/15)^{3/2}]1.1\times
10^{45}\left[X^{1+\alpha}Z^{2}F_{151}\right]^{0.857}\mathrm{ergs/s}\;,\\
&& Z \equiv 3.31-(3.65)\times\nonumber \\
&&\left[X^{4}-0.203X^{3}+0.749X^{2} +0.444X+0.205\right]^{-0.125}\;,
\end{eqnarray}
where $X\equiv 1+z$, $F_{151}$ is the total optically thin flux
density from the lobes (i.e., contributions from Doppler boosted
jets or radio cores are removed). This sophisticated calculation of
the jet kinetic luminosity incorporates deviations from the overly
simplified minimum energy estimates into a multiplicative factor
\textbf{f} that represents the small departures from minimum energy,
geometric effects, filling factors, protonic contributions and low
frequency cutoff \cite{wil99}. The quantity, \textbf{f}, was further
determined to most likely  be in the range of 10 to 20 \cite{blu00}.
In this paper we adopt the following cosmological parameters:
$H_{0}$=70 km/s/Mpc, $\Omega_{\Lambda}=0.7$ and $\Omega_{m}=0.3$.
Define the radio spectral index, $\alpha$, as
$F_{\nu}\propto\nu^{-\alpha}$. The formula is most accurate for
large classical double radio sources, thus we do not consider
sources with a linear size of less than 20 kpc which are constrained
by the ambient pressure of the host galaxy. Alternatively, one can
also use the independently derived isotropic estimator in which the
lobe energy is primarily inertial in form \citet{pun05}
\begin{eqnarray}
&&\overline{Q}\approx
5.7\times10^{44}(1+z)^{1+\alpha}Z^{2}F_{151}\,\mathrm{ergs/sec}\;.
\end{eqnarray}
Due to Doppler boosting on kpc scales, core dominated sources with a
very bright one sided jet (such as 3C 279 and most blazars) must be
treated with care \citep{pun95}. The best estimate is to take the
lobe flux density on the counter-jet side and multiply this value by
2 (bilateral symmetry assumption) and use this estimate for the flux
density in Equations (6) - (8).
\section{Sample Selection} Determination of quasar EUV continua requires space based
observations of modest redshift quasars since ground based
observations of high redshift objects in which the EUV is redshifted
into the optical are heavily attenuated by the Ly$\alpha$ forrest
\citep{zhe97,tel02}. In order to get a meaningful estimate of
$\alpha_{\mathrm{EUV}}$, a range of at least 700 \AA\, to 1100 \AA\,
in the quasar rest frame is needed to extract the continuum from the
numerous broad emission lines \citep{tel02}. Therefore, a redshift
of $z>0.63$ is required. Troughs from Lyman limit systems (LLS) were
removed by assuming a single cloud with a $\nu^{-3}$ opacity. This
was considered acceptable if the power law above the LLS could be
continued smoothly through the corrected region (see the spectra in
the Appendix). If there were many strong absorption systems or a LLS
that compromised a broad emission line, this simple procedure was
deemed inadequate for continuum extraction with the available data
and the spectrum was eliminated from the sample. A small correction
for the Lyman valley was also made \citep{zhe97}. Additionally, if
there was evidence of a blazar synchrotron component contribution to
the continuum such as a dominant flat spectrum radio core
accompanied by high optical polarization or optical/UV variability,
or low equivalent width of the emission lines, the underlying
accretion disk continuum was considered too uncertain for the
sample.
\begin{table*}
 \centering
\caption{The Jet Power of the HST RLQ Sample} {\small
\begin{tabular}{ccccccc}
 \hline

 Source & Alias & z & Lobe Flux Density & Observed Frequency  & Estimated $Q$  \\
        &       &   & (mJy)   & (GHz)   &  ($10^{45}$ ergs/s) \\
\hline
0024+224  &  ... & 1.12 & 11   & 1.4    & $1.38 \pm 0.67$\\
0232-042  &  PKS 0232-04 & 1.45 & 6100 & 0.178 & $32.45 \pm 8.81$\\
0637 -7516 & PKS 0637 -752 & 0.65 & 209  & 4.8 & $2.83 \pm 1.24 $\\
0743-67  &  PKS 0743-67 & 1.51 & 1450   & 2.5 & $65.87\pm 14.4$\\
0959+6827 &  ... & 0.77 & 27   & 1.4 & $ 0.27 \pm 0.16 $\\
1022+194 &  4C +19.34 & 0.83 & 2400   & 0.178 & $4.01 \pm 1.67$\\
1040+123 &  3C 245 & 1.03 & 1179   & 1.4 & $15.18\pm 4.95$\\
1137+660 &  3C 263 & 0.65 & 18380   & 0.151 & $13.17 \pm 4.42 $\\
1229-021 &  4C -02.55 & 1.05 & 9000  & 0.160 & $20.52 \pm 6.25 $\\
1241+176 &  PG 1241+176  & 1.27 & 112   & 1.4 & $2.90 \pm 1.27 $\\
1244+324  &  4C +32.41 & 0.95 & 3370   & 0.151 & $6.53\pm 2.51$\\
1252+119 &  PKS 1252+119 & 0.87 & 15   & 1.4 & $0.31\pm 0.18$\\
1317+5203 &  4C+57.21 & 1.06 & 3800   & 0.178 & $10.38 \pm 3.66 $\\
1340+289 &  FBQS J1343+2844 & 0.91 & 830   & 0.151 & $1.67 \pm 0.79$\\
1340+606 &  3C 288.1 & 0.96 & 9900   & 0.151 & $17.83 \pm 5.61$\\
1354+19 &  PKS 1354+19  & 0.72 & 545   & 1.4 & $3.57\pm 1.32$\\
1415+172 &  PKS 1415+172  & 0.82 & 1020   & 0.408 & $3.31\pm 1.42$\\
1857+566 &  4C +56.28 & 1.60 & 8650  & 0.160 & $48.99 \pm 11.82$\\
2149+212 &  4C +21.59 & 1.54 & 5300   & 0.160 & $30.20\pm 8.35 $\\
2340-036 &  PKS 2340-036  & 0.90 & 61   & 1.4 & $ 0.78\pm 0.41$\\

\hline
\end{tabular}}
\end{table*}
\begin{table*}
 \centering
\caption{The EUV Properties of the HST RLQ Sample} {\tiny
\begin{tabular}{cccccccc}
 \hline

 Source & Alias & z & $\lambda L_{\lambda}(\lambda = 1100\AA)$ & $L_{\mathrm{bol}}$ & $\alpha_{EUV}$   & Spectrograph/   \\
        &       &   & ($10^{45}$ ergs/s)& ($10^{45}$ ergs/s)&  $         $      &  Grating     \\
\hline
0024+224  &  ... & 1.12 & 64,05   & 243.50 & $1.77 \pm 0.09 $& FOS/G160L  \\
0232-042  &  PKS 0232-04 & 1.45 & 94.47 & 358.98 & $ 1.75\pm 0.09 $ & FOS/G160L, G270H \\
0637 -7516 & PKS 0637 -752 & 0.65 & 20.80  & 79.02 & $ 1.75\pm 0.09 $& FOS/G160L \\
0743-67  &  PKS 0743-67 & 1.51 & 110.78  & 420.95 & $ 2.20 \pm 0.30 $& FOS/G190H, G270H\\
0959+6827 &  ... & 0.77 & 22.28   & 84.65 & $ 1.40 \pm 0.10 $& FOS/G160L \\
1022+194 &  4C +19.34 & 0.83 & 7.02   & 26.67 & $ 2.55\pm 0.15 $ & FOS/G160L  \\
1040+123 &  3C 245 & 1.03 & 11.90   & 45.21 & $2.20\pm 0.20 $& FOS/G160L \\
1137+660 &  3C 263 & 0.65 & 27.39   & 104.10 & $ 2.00\pm 0.10$ & COS/G130M, G160M; FOS/G190H\\
1229-021 &  4C -02.55 & 1.05 & 25.09  & 95.36 & $ 2.65\pm 0.15$ &  FOS/G160L, G190H, G270H\\
1241+176 &  PG 1241+176  & 1.27 & 86.78   & 329.76 & $ 1.79\pm 0.09$ & STIS/G230L \\
1244+324  &  4C +32.41 & 0.95 & 13.77   & 52.31 & $ 2.41\pm 0.25 $ & FOS/G160L  \\
1252+119 &  PKS 1252+119 & 0.87 & 16.12   & 61.27 & $1.45\pm 0.10$ &  FOS/G160L, G190H\\
1317+5203 &  4C+57.21 & 1.06 & 37.43   & 142.25 & $2.18\pm 0.20 $ & FOS/G160L   \\
1340+289 &  FBQS J1343+2844 & 0.91 & 14.76   & 56.08 & $190 \pm 0.10 $ & FOS/G160L    \\
1340+606 &  3C 288.1 & 0.96 & 11.39   & 43.27 & $1.78 \pm 0.09 $ &  STIS/G140L, G230L \\
1354+19 &  PKS 1354+19  & 0.72 & 27.46   & 104.34 & $1.95\pm 0.10 $ &  FOS/G160L  \\
1415+172 &  PKS 1415+172  & 0.82 & 7.28   & 27.66 & $ 1.93 \pm 0.10 $ & FOS/G160L   \\
1857+566 &  4C +56.28 & 1.60 & 22.60  & 85.87 & $ 2.87\pm 0.14 $ & STIS/G230L \\
2149+212 &  4C +21.59 & 1.54 & 18.22   & 69.24 & $ 2.27\pm 0.20 $ &  STIS/G230L \\
2340-036 &  PKS 2340-036  & 0.90 & 38.78   & 147.35  & $ 1.88\pm 0.09$ & FOS/G160L  \\

\hline
\end{tabular}}
\end{table*}
\par As discussed in the last section, the most reliable methods of estimating the long
term time averaged jet power $Q$ are based on the the optically thin
emission from relaxed radio lobes. Thus, all sources in the sample
needed proof of extended emission on scales larger than the host
galaxy so that the lobes can relax ($> 20$ kpc). Verification
required archival high resolution interferometry images made between
0.408 GHz and 5 GHz. The HST and radio selection criteria resulted
in a total of 20 sources for the sample. Note that two new southern
hemisphere sources have been added to the sample of spectra from
\citet{pun14}. The Notes on Individual Sources in the Appendix
provides the details that allow these sources to pass the criteria
required to be in the sample. The optically thin emission was
estimated based on 151 MHz - 178 MHz flux densities (if available)
and the lobe fluxes from the radio images. The largest spread in the
estimates of $Q$, based on optically thin extended emission, are
bounded on the high side by Equation (6) for the parameter,
\textbf{f} = 20, and on the low side by Equation (8). These two
extremes are used to generate the uncertainty in $Q$ in Table 1.
\begin{figure*}
\includegraphics[width=170 mm, angle= 0]{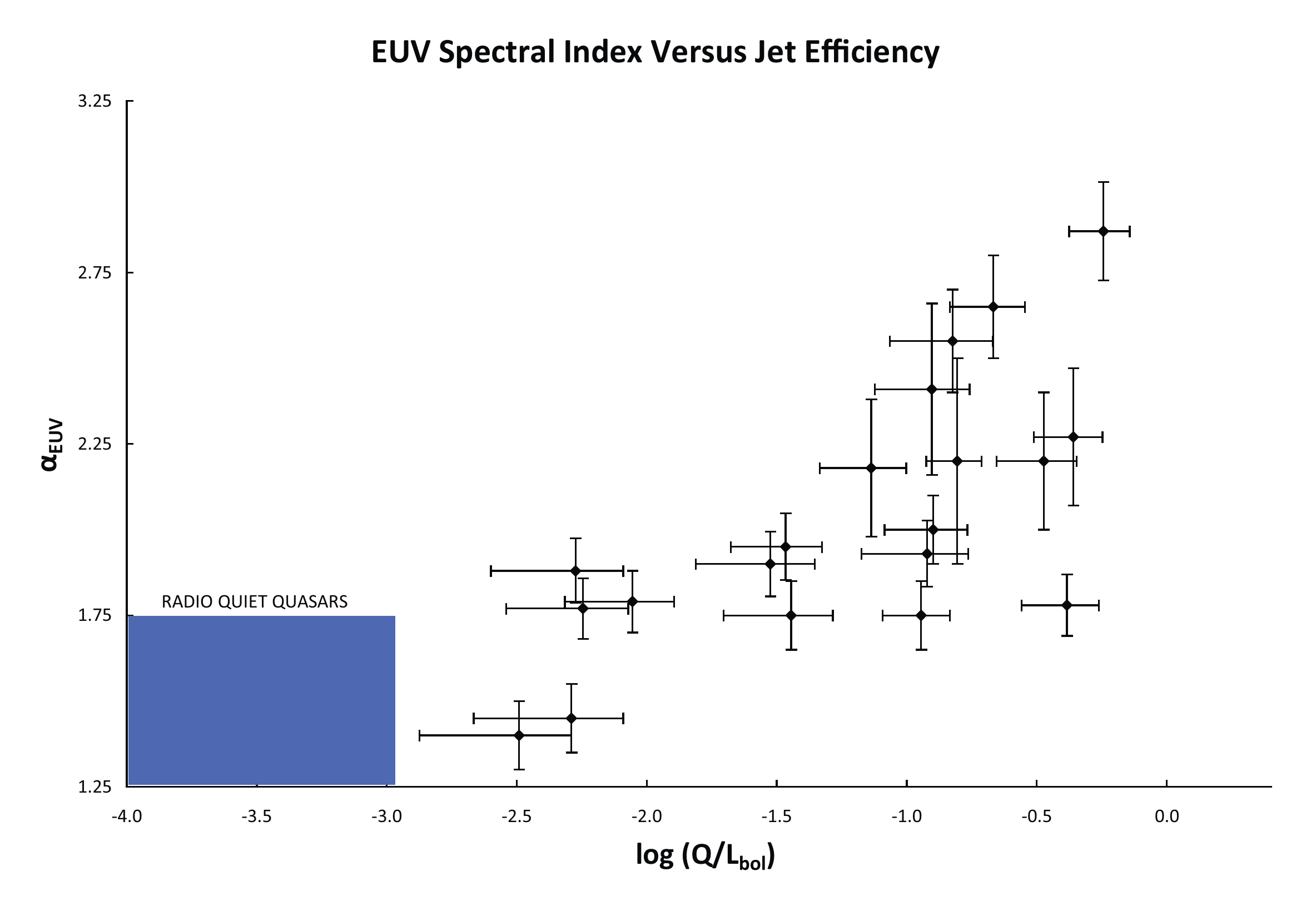}
\caption{The EUV Deficit versus the Jet Efficiency. A scatter plot
of $Q/L_{\mathrm{bol}}$ and $\alpha_{\mathrm{EUV}}$. The blue
rectangle represents the RQQs for comparison. The RQQs composite of
Figure 1 has $\alpha_{\mathrm{EUV}} = 1.57\pm 0.17$ (Telfer et al
2002). Another composite of predominantly RQQs has
$\alpha_{\mathrm{EUV}}=1.41\pm 0.16$ (Stevans et al 2014). So a
broad range of $1.50\pm 0.25$ is chosen to span both composites.}
\end{figure*}
\begin{figure*}
\includegraphics[width=170 mm, angle= 0]{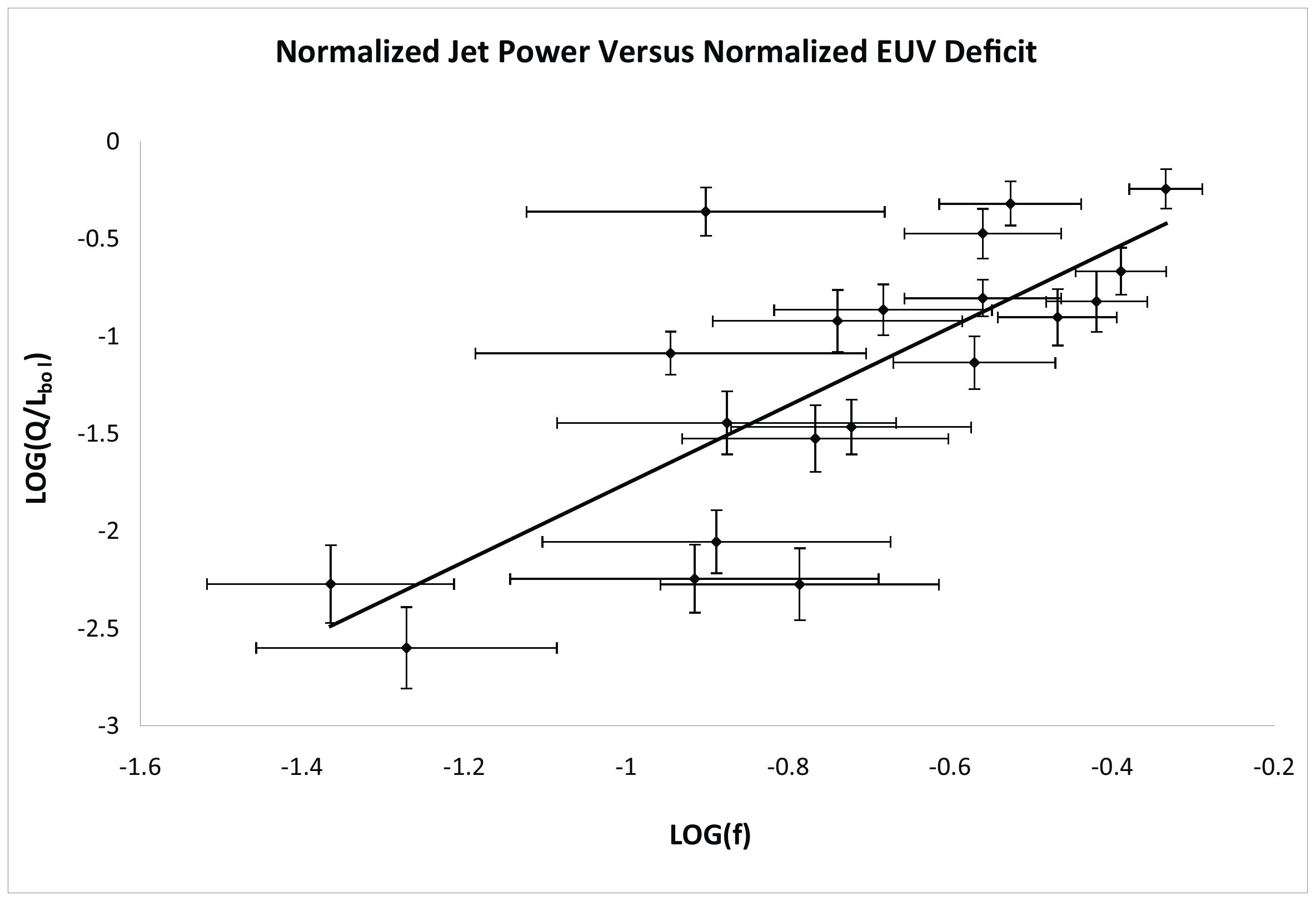}
\caption{Jet Power and the EUV Deficit. The plot of
$\log[Q/L_{\mathrm{bol}}]$, normalized jet power, versus normalized
EUV deficit, $\log(f)$. The best fit line is plotted in black for
the case that the lower cutoff on $f$ is 0.025. The slope is 2.005
in agreement with the exponent of 2 in Equation (11). The
determination of $f$ for small values (flat EUV spectrum sources) is
very sensitive to the choice of the fiducial radio quiet EUV level,
hence the large error bars for small $f$.}
\end{figure*}
\section{Experiment Design and Results}
Equations (4) and (5) are two approximate basic and testable
predictions of the MAA explanation for the EUV deficit in RLQs. In
order to design an experiment to test the validity of the relations,
one needs a sample of RLQs with adequate EUV spectra and radio
observations to perform the test as described in the last section.
Equations (4) and (5) are tested by looking at the HST spectra that
cover the span from the SED peak at 1100 \AA\, to 700 \AA\,. The
first experiment tests Equation (4) by assessing the implied
correlation between $Q/L_{\mathrm{bol}}$ and
$\alpha_{\mathrm{EUV}}$, where the spectral luminosity of the
continuum is approximated as $L_\nu \propto
\nu^{-\alpha_{\mathrm{EUV}}}$. Since the data used here covers the
peak of the SED at $\lambda \approx1 100 \AA$, an accurate
expression, $L_{\mathrm{bol}} \approx 3.8 \lambda L_{\lambda}
(\lambda =1100 \AA)$, can be used to estimate the accretion disk
luminosity proper (less reprocessed IR emission in distant molecular
clouds) \citep{dav11,pun14}. The relevant data is captured in Tables
1 and 2. The data scatter is plotted in Figure 3. If the MAA
scenario is correct not only would there be a correlation between
$\alpha_{\mathrm{EUV}}$ with $Q/L_{\mathrm{bol}}$, but the
correlation of $\alpha_{\mathrm{EUV}}$ with $Q/L_{\mathrm{bol}}$
should be stronger than $\alpha_{\mathrm{EUV}}$ with $Q$ since
dividing by $L_{\mathrm{bol}}$ is equivalent to dividing out the
scatter induced by the $B_{z}^{2}$ term in Equation (3). The
Spearman rank correlation test indicates that the probability that
the correlation of $\alpha_{\mathrm{EUV}}$ and $Q/L_{\mathrm{bol}}$
in Figure 3 occurs by random chance is 0.001. Note this is a more
significant correlation compared to a probability of the scatter
occurring by random chance of 0.006 for $\alpha_{\mathrm{EUV}}$ and
$Q$ and 0.150 for $\alpha_{\mathrm{EUV}}$ and redshift (z). Thus,
the expected correlation exists. Physically, the stronger
correlation of $Q/L_{\mathrm{bol}}$ with $\alpha_{\mathrm{EUV}}$
compared to $\alpha_{\mathrm{EUV}}$ and $Q$ is evidence that the
poloidal magnetic pressure in the islands is being set by the ram
pressure of the surrounding accretion flow. The correlation of
$Q/L_{\mathrm{bol}}$ with $\alpha_{\mathrm{EUV}}$ is definitely
improved, but not dramatically improved from that with $Q$ and
$\alpha_{\mathrm{EUV}}$. This is likely a consequence of the
circumstance that the correlation with $Q$ is already very strong.
Allowing for cosmic scatter generating effects (i.e., a simplified
uniform MAA assumption, modest geometric variations from object to
object and epoch to epoch within the same object, plus
non-contemporaneous measurement of $Q$ with $L(EUV)$), it is
probably not realistic to expect that the probability that the
correlation of $Q/L_{\mathrm{bol}}$ with $\alpha_{\mathrm{EUV}}$
occurs by random chance is 0 to more than 2 significant digits. In
order to see if this is a statistically significant result consider
the partial correlation of $Q/L_{\mathrm{bol}}$ with
$\alpha_{\mathrm{EUV}}$ when $Q$ is held fixed. The partial
correlation coefficient is 0.492 which correspondence to a
statistical significance of 0.984. Conversely, the partial
correlation of $Q$ with $\alpha_{\mathrm{EUV}}$ when
$Q/L_{\mathrm{bol}}$ is held fixed is significant at the 0.581
level. Similarly, repeating this analysis with the Kendall tau rank
test, yields: the partial correlation of $Q/L_{\mathrm{bol}}$ with
$\alpha_{\mathrm{EUV}}$ when $Q$ is held fixed is statistically
significant at the 0.957 level and the partial correlation of $Q$
with $\alpha_{\mathrm{EUV}}$ when $Q/L_{\mathrm{bol}}$ is held fixed
is statistically significant at the 0.753 level. There is
statistical evidence that the correlation of $Q/L_{\mathrm{bol}}$
with $\alpha_{\mathrm{EUV}}$ is the fundamental correlation and the
correlation of $Q$ with $\alpha_{\mathrm{EUV}}$ is a spurious
correlation. This supports the premise of equation (3) that it is
the jet power divided by accretion rate (ram pressure) that is the
physical variable that is related to the EUV deficit. Finally, the
much weaker correlation with redshift indicates that more profound
physics is occurring, not just cosmic evolution or selection
effects.

\par The second experiment is to test Equation (5) which is much more restrictive than Equation (4) and much more
sensitive to any initial assumptions and deviations from the
homogeneous approximation. In particular, this experiment will test
both the $Q/L_{\mathrm{bol}}$ dependence on $f^{2}$ and the
connection between $f$ and the EUV deficit in Equation (1). First,
define $L(\mathrm{EUV})$ in a normalized form since from Equation
(1) it scales with $SA$. The shortest wavelength that is uniformly
sampled is $\lambda = 700 \AA$, so $L(\mathrm{EUV}) \propto
L_{\nu}(\lambda= 700\AA)$ is the best available proxy for
$L(\mathrm{EUV})$. In order to remove the dependence on $SA$ this
value is normalized by the peak of the SED
\begin{equation}
L(\mathrm{EUV}) \equiv \frac{L_{\nu}(\lambda=
700\AA)}{L_{\nu}(\lambda= 1100\AA)} \;.
\end{equation}
For RQQs, the average value of $\alpha_{\mathrm{EUV}}= 1.57 \pm
0.17$ from \citet{tel02}. As noted in Figure 3, another composite of
predominantly intermediate redshift RQQs has
$\alpha_{\mathrm{EUV}}=1.41\pm 0.16$ \citep{ste14}. Thus, there is
no unique estimate of the fiducial or baseline EUV level for the
absence of magnetic islands. The variation in the RQQ EUV translates
into an uncertainty in the estimate for $f$ in Equation (5). Denote
the maximum and minimum estimates of the EUV in the absence of
magnetic islands determined from the HST composite spectral analysis
as
\begin{eqnarray}
&& L(\mathrm{EUV})_{\mathrm{RQQ}}\biggr\vert_{\mathrm{max}} \equiv
\frac{L_{\nu}(\lambda= 700\AA)}{L_{\nu}(\lambda= 1100\AA)}
\biggr\vert_{\alpha=1.25} = 0.57 \;, \\ \nonumber
&&
L(\mathrm{EUV})_{\mathrm{RQQ}}\biggr\vert_{\mathrm{min}} \equiv
\frac{L_{\nu}(\lambda= 700\AA)}{L_{\nu}(\lambda= 1100\AA)}
\biggr\vert_{\alpha=1.74} = 0.46 \;.
\end{eqnarray}
Combining Equations (5), (9) and (10) yields the crude, but simple
prediction of the homogeneous MAA model applied to these
intermediate redshift quasars observed with HST,
\begin{equation}
Q/L_{\mathrm{bol}} \approx Af^{2}\approx A\left[1 -
\left[\frac{L_{\nu}(\lambda= 700\AA)}{L_{\nu}(\lambda=
1100\AA)}\right]\left[\frac{L_{\nu}(\lambda=
1100\AA)}{L_{\nu}(\lambda=
700\AA)}\right]\biggr\vert_{\mathrm{RQQ}}\right]^{2} \;.
\end{equation}
The larger the estimate for the RQQ EUV baseline level in Equation
(10), the larger the value of $f$ (the EUV deficit) that is computed
in Equation (11). Equation (11) is utilized as follows in Figure 4
for estimating $f$: the expression for $f$ in Equation (11) is
computed two ways, one with the maximum baseline EUV luminosity from
Equation (10) and again with the minimum baseline luminosity in
Equation (10). The results of these two values in the expression for
$f$ in Equation (11) are averaged, this is the $f$ value that is
plotted in Figure 4. The uncertainty in this average $f$ is
calculated as the maximum of the the expression for $f$ in Equation
(11) minus the average. Implementing Equation (10) in Equation (11)
is not completely straightforward since according to Table 2, two of
the sources have $\alpha_{\mathrm{EUV}} < 1.74$ and two others
quasars have $\alpha_{\mathrm{EUV}}$ that are close to this value.
The volume of the EUV emitting region that is displaced by magnetic
islands cannot be negative, $f$ cannot be less than 0 in Equation
(11). Thus, consider the example of a lower bound or cutoff of at
least $2.5\%$ of the accretion surface area in the EUV region ($f >
0.025$) is displaced by magnetic islands for the RLQs with flatter
EUV spectrum when the upper limit value of $\alpha_{\mathrm{EUV}} =
1.74$ is used in Equation (11) as the condition for the absence of
magnetic islands. The result is shown graphically in Figure 4 for
each source in Tables 1 and 2. This cutoff is arbitrary, so the
effect of its variation is investigated below.
\par The second experiment in Figure 4 plots the logarithm of $Q/L_{\mathrm{bol}}$
versus logarithm of $f$ from Equation (11) from the data in Tables 1
and 2. The black line is the linear fit to these 20 sources and the
slope is 2.01. The theoretical prediction of MAA is 2 in Equation
(11). The incredible agreement is clearly coincidental since there
is significant scatter and many crude approximations in the
derivation of Equation (11). As an illustration of this claim,
consider a lower bound of $f>0.05$. In this case, the slope is 2.22.
Alternatively, if the lower bound of $f>0.01$ is used, the slope is
1.88. The fit to the best fit line in Figure 4 is fairly good with a
squared multiple regression correlation coefficient of 0.57.

\section{Implications and Conclusion}
The primary results of this study that are presented in Tables 1 and
2 (as well as Figures 3 and 4) are empirical. It was shown that HST
EUV spectra and radio interferometry data are consistent with the
jet power, $Q$, in RLQs proportional to both the square of the
deficit in EUV emission, $f^{2}$, and the accretion rate. The most
straightforward interpretation is to assume that magnetic flux tubes
occupy a fraction, $f$, of the EUV emitting region since this
explains the EUV deficit and jet power connection. Furthermore, it was argued
that the MAA/EMAA variants of this idea explains the precise scaling laws that are
implied by the observational data. In the following, the experiments designed to test
the MAA hypothesis are summarized. The speculative nature of the MAA
scenario is critiqued as well as alternative explanations of the
observed scaling laws.

\par In the first experiment, the scatter in Figure 3 indicates that
$Q/L_{\mathrm{bol}}$ is strongly correlated with
$\alpha_{\mathrm{EUV}}$ as expected from Equation (4), i.e., the jet
launching region displaces the EUV emitting region in MAA. Secondly,
it was shown that the correlation of $Q/L_{\mathrm{bol}}$ with
$\alpha_{\mathrm{EUV}}$ is stronger than the correlation of $Q$ with
$\alpha_{\mathrm{EUV}}$. It was argued that this is a statistically
significant difference because a partial correlation analysis
indicates that the correlation of $Q/L_{\mathrm{bol}}$ with
$\alpha_{\mathrm{EUV}}$ is statistically significant and the
correlation of $Q$ with $\alpha_{\mathrm{EUV}}$ is spurious. This is
evidence of the precise details of MAA that is elucidated by
Equation (3), ram pressure regulates the strength of the large scale
poloidal field strength in the magnetic islands.

\par In the second experiment, Figure 4 shows a data scatter that is consistent with the jet strength
being proportional to the square of the EUV deficit estimated from
HST spectra in RLQs. The fitted slope of the log-log relationships
between $Q/L_{\mathrm{bol}}$ and $f$ (Figure 4 is plotted for a
lower cutoff $f=0.025$) are consistent with the prediction of
Equation (11) of a slope equal to 2. Namely with reasonable
variation due to the lower cutoff ($0.01<f<0.05$) the fitted slope
was found to be $2.05 \pm 0.17$. This is significant evidence that
the fundamental details of the crude MAA model are consistent with
the observations. Figure 4 indicates that $f$ is a few percent for
the RLQs with weaker jets and $\sim 40\%$ for the RLQs with the most
powerful jets.

\par It is not claimed that this is the only possible explanation of
the EUV deficit in RLQs. Other explanations based on numerical and
theoretical models include, lower $a$ in RLQs (larger innermost
stable orbit), or stronger quenching winds in RLQs per the model of
\cite{lao14}. Note that a scenario of larger black hole mass and
lower accretion rates in RLQs was ruled out empirically as a
plausible explanation due to the indistinguishable SED peak in RLQs
and RQQs \citep{pun14}. However, none of these other explanations
naturally produces a radio jet and the correlation in Figure 3.

\par Furthermore, it is also not claimed that this is the only possible explanation of $Q$
being controlled by $P_{\mathrm{ram}}$.  Magnetic flux pinned to the
event horizon as in \citet{mck12,tch11} or trapped in an equatorial
gap between the accretion flow and the black hole as in
\citet{pun08} would also result in this same correlation. However,
neither of these scenarios explain the $Q/L_{\mathrm{bol}}$ scaling
with the square of the EUV deficit, $f^{2}$ found in Figure 4. A
possible reconciliation is provided near the end of this section.

\par In Section 2, it was noted that ``if the magnetic islands are not
extremely short-lived transient features in the inner accretion
flow, an approximate balance of the ram pressure and the magnetic
pressure of the poloidal magnetic field in the islands must exist."
Currently, the diffusion rate of plasma onto and off of magnetic
field lines and magnetic reconnection rates are not well known near
black holes. Not only are these issues critical for the formation of
the magnetic islands, but the time evolution of the magnetic islands
is determined primarily by diffusion \citep{igu08,pun15}. These
dynamical elements only occur as a consequence numerical diffusion
in modern simulations in over simplified ideal MHD single fluid
models of the physics \citep{pun15}. It is not even clear
theoretically what the basic principles required for an accurate
physical depiction would be. Many issues that are related to these
topics are active areas of investigation in solar and fusion physics
\citep{bau13,mal09,thr12,yam07}.  As such, it cannot currently be
claimed or refuted that long-lived magnetic islands can or cannot
exist in the inner accretion flow. The magnetic islands should
become Kruskal- Schwarzschild unstable if the density becomes low
enough and should move outward, not inward. A key physical element
is the numerical approximation of the physics that determines the
density. If the numerical diffusion is significant, low enough
density may not be achieved before the flux tubes are dragged across
the event horizon. The simulations in \citet{igu08,pun09} indicate a
population of magnetic islands within the innermost accretion flow
consistent with the range of filling factors, $f$, that are found in
Figure 4. In the innermost accretion flow, the time evolution of the
islands is non-steady. The inner accretion flow is a dynamic region.
There are epochs with very few magnetic islands and epochs in which
the (large) magnetic islands near the black hole become buoyant
after losing mass to the jet and slowly wind their way out against
the ram pressure of the accretion flow. By contrast, the simulations
in \citet{mck12,tch11,tch12} that are heavily seeded with large
scale magnetic flux are devoid of magnetic islands close to the
event horizon. This is evidenced by the claim in \citet{mck12} that
no significant Poynting flux emerges from this region (see Equation
2, above) as well as the linked online videos of the simulations.
The videos show the innermost significant, modest, magnetic island
concentrations are located at $r > 10M$ and they are extremely
transient. Thus, even though the simulations in
\citet{mck12,tch11,tch12} have strong Poynting jets from the event
horizon, they are not consistent with the EMAA model. This either
means that the interpretation of the EUV deficit presented here is
wrong or the simulations do not represent the magnetic flux
evolution accurately. The latter option is favored empirically,
since magnetic flux displacing EUV emitting gas in the inner
accretion flow explains two correlated phenomena, jet power and the
EUV deficit. It not only explains the correlation, but it predicts
the explicit scaling laws for jet power with accretion rate and with
the degree of EUV suppression implied by the observations. The
simulations without significant vertical magnetic flux in the
innermost accretion flow explain only the scaling law for jet power
with accretion rate.

\par Recall the fact that the
only difference between the RLQ and RQQ composite continua in Figure
1 is the EUV. Comparing this difference to numerical simulations
indicates that the magnetic islands are concentrated between the
event horizon and an outer boundary of $<2.8 M$ (in geometrized
units) for rapidly rotating black holes \citep{pun14,pen10}. Thus
the simple, homogeneous MAA model does not preclude a black hole
spin assisted component that displaces the EUV emitting gas inside
$r < 2M$ as in the ''erogspheric disk" that has been found to occur
in some 3-D numerical simulations of high spin black holes
\citep{pun08,pun09}. It is also possible that a certain fraction of
the flux distribution in the inner accretion flow threads the event
horizon and extracts spin energy \citep{bla77}. In particular, what
is shown here is that a significant magnetic flux in the inner
accretion flow (based on Figure 4 from 2.5\% to 45\% of the EUV
emitting region in RLQs is filled with magnetic islands) explains
the EUV deficit and the source of the jet power is proportional to
the square of this putative flux. However, the distribution of flux
can be larger than just the inner accretion flow. For example, the
magnetic flux distribution might in general thread the horizon and
the ergospheric equatorial plane as well. This is the EMAA model
that was sketched out in the introduction. The magnetic islands in
the innermost accretion flow would represent the outermost portion
of this flux distribution in this scenario. This inner region of
magnetic flux would be proportional to the flux that threads the EUV
region for the putative generic flux distribution,
$\Phi_{int}\propto \Phi_{EUV}$. The power from the event horizon jet
and/or the ergospheric disk jet could be larger than the jet power
from the EUV region. In this scenario, the flux in the innermost
accretion flow is merely a ``tracer" for flux contained in these
interior regions and not coincident with the primary source of the
Poynting flux that powers the jet. However, the flux permeating the
innermost accretion flow must be significant and the flux in these
interior regions is presumed to scale with that in the EUV region.
Namely, from Figure 4, for a weak jet like that in PKS 1252+119,
$f\sim 4\%$ and for a strong jet like that of 1857+566 $f\sim 45\%$.
By the EMAA assumption, one would expect the event
horizon/ergosphere flux to be $\sim 10$ times larger in 1857+566
compared to PKS 1252+119 on average (and therefore
$Q/L_{\mathrm{bol}}$ $\sim 100$ times larger) in order to explain
the observational results (Figures 3 and 4) that are described in
this paper. The implication is still that simulations of jets from
these regions would need to have a significant flux in the innermost
accretion flow to be consistent with the observations.  As with the
discussion of the time evolution of magnetic islands above, the
existing numerical algorithms are single fluid MHD approximations to
the physics in which numerical diffusion determines reconnection
rates and diffusion rates and may not be reliable depictions of the
physical model of magnetic field dynamics in the innermost accretion
flow near black holes. As such, a different mathematical description
of the reconnection process can lead to very different poloidal
magnetic flux distributions near the black hole \citep{pun15}.
Hence, much of this discussion is speculative. On a less speculative
note, the fundamental deduction drawn from the EUV observations is
that future modeling and theory of quasar jet origins should contain
the feature of an innermost accretion flow threaded by substantial
vertical magnetic flux.
\par It is necessary for a critical analysis that one segregates
the various issues addressed in the MAA/EMAA model and the numerical
simulations by the degree to which they have been verified by the
scientific method, observation. The hierarchal list below is in order of the
degree to which each of these issues is verified by observations.
\begin{enumerate}

\item EUV deficit and radio loudness: The EUV deficit is
verified observationally here and in \citet{pun14} to correlate with
low frequency radio luminosity on super-galactic scales (converted
to jet power here in order to make contact with theoretical
treatments).

\item EUV originates from inner accretion flow: Being on the
high frequency tail of the optically thick thermal spectrum this is
the obvious interpretation. However, observational verification is
scientifically much more robust. The only direct method of
estimating the size of the EUV region is through time variability
arguments. This requires the ability to collect a significant EUV
flux on short time scales. The only reasonable set of observations
related to an active nucleus are the EUVE (Extreme Ultraviolet
Explorer) observations of NGC 5548. Simultaneous UV and EUV
monitoring indicated that the most likely interpretation of the EUV
emission was the Wien tail of the optically thick thermal emission
and it had significant variability \citep{mar97}. Further EUV
monitoring found significant variability at the smallest sampling
time scales $t_{var} < 5600$ s \citep{hab03}. Standard arguments
based on the light travel time across the EUV emitting region and
black hole reverberation mass estimates of the central black hole
indicate the EUV emitting gas is located in a volume with a radius,
$r<10.5$M \citep{ben07,den10}. The light curve in \citet{hab03}
seems to indicate that with a higher sensitivity telescope, the time
scale for variability would likely be less than 5600 s. This is
observational evidence that the EUV is radiated from the innermost
accretion flow.

\item The only know energy source for the extreme powers in relativistic quasar
jets is Poynting flux (independent of the details of its origins)
and this scales with the square of the enclosed poloidal magnetic
flux.

\item Magnetically arrested accretion occurs in some numerical
models, however the dynamics of the arresting magnetic flux tubes
differs from what is indicated in the MAA/EMAA model to varying
degrees. The difference is small to modest in \citet{igu08,pun09},
that agree with the fill fraction, $f$, expected from Figure 4 in
the innermost accretion flow. The time evolution generally agrees
with the MAA/EMAA model except that some of the islands may not
agree due to some very unstable magnetic outbursts at the smallest
radii. The difference from the MAA/EMAA explanation of the
observations is large in \citet{mck12,tch11,tch12} for which fill
fraction, $f\approx 0$ in the innermost accretion flow. The
``magnetically choked accretion flows" of \citet{mck12} heat the
innermost accretion flow compressively as part of the magnetic
choking process. This likely eliminates the EUV region by
drastically elevating the temperature of the innermost accretion
flow as opposed to suppressing the emission from the innermost
accretion flow as the observations indicate. Alternatively,
depending on the precise, unknown, details of radiative transfer in
this hot, dense gas, the compressive heating might drastically
harden the high frequency tail of the optically thick spectrum in
RLQs, the opposite of what is observed.

\end{enumerate}
This list is the basis of the logic of the analysis presented in
this study. Items 1 and 2 above state that observations directly
indicate that jets disrupt (suppress) the EUV emission from the
inner accretion disk and the amount of disruption scales with the
power of the jet. The third point states that the only known
mechanism for driving such a powerful relativistic jet is Poynting
flux that requires significant poloidal magnetic flux at its source.
Thus, with very little speculation, it is indicated that both the
large scale poloidal magnetic flux at the base of the jet and a
mechanism that suppresses (but does not eliminate) the EUV emission
from the innermost accretion flow coexist at the heart of the
central engine of RLQs. The main assumption of the MAA/EMAA idea is
that this is too large of a coincidence, some of this magnetic flux
must be the same element that disrupts, but does not eliminate the
innermost accretion flow. If a numerical effort cannot reproduce
this circumstance then perhaps the numerical approximation to the
relevant physical processes requires further development.
Observations should lead the numerical work.

\begin{acknowledgements}
I would like to thank Michael Shull and
 Matt Stevans for sharing their latest spectral fit to 3C 263. I would also like to thank Robert Antonucci
 for his valuable critique of a previous version of the manuscript. This research was supported by
 ICRANet. The research benefitted from the very knowledgable referee who greatly improved the manuscript.
\end{acknowledgements}

\section*{Appendix 1.: Notes on Individual Sources}
\par 0024+224. The NVSS image at 1.4 GHz shows a predominantly
core-jet morphology.  The one-sided jet is likely Doppler boosted
kpc emission as is typical of most RLQs viewed close to the jet axis
\cite{pun95}. There are two faint flux density peaks just beyond the
hot spot at the end of the jet. This is taken as evidence of
isotropic halo type emission and this flux density is used in the
estimate of Q. Even though the radio source is core-dominated, the
optical polarization and optical variability are small and the
emission lines have high equivalent widths, Thus, Doppler boosted
jet emission in the EUV is considered negligible.

0232-042. This is a strong lobe dominated source, so the low
frequency total flux density at 178 MHz is the best estimator for Q.

0637-752. This is a famous core plus one-sided kpc jet dominated
southern hemisphere radio source. There is clearly considerable
Doppler beaming. The most reliable estimate of isotropic flux is to
take twice the counter lobe flux density as discussed above. This is
attained from the ATCA 4.8 GHz radio image \cite{tin98}. Even though
the radio source is core-dominated, the optical polarization and
variability are small and the emission lines have high equivalent
widths, Thus, Doppler boosted jet emission in the EUV is considered
negligible.

0743-67. This is one of the most intrinsically luminous quasars and
is also a very strong radio source, both the radio core and the
lobes. The lobe flux density is estimated from ATCA radio
observations \cite{tin05}.  Even though the radio core exceeds 1 Jy,
the optical polarization and variability are small and the emission
lines have high equivalent widths, Thus, Doppler boosted jet
emission in the EUV is considered negligible. This source has
significantly higher visual extinction than any other source in the
sample, $A_{V}$ = 0.362. Thus, unlike the other sources, the
detailed form of the Galactic extinction law significantly affects
the EUV continuum fit. The average Galactic value of $R_{V}$ = 3.1,
gives a poor fit to a power-law continuum for the CCM model
\cite{car89}. The reason is clear, de-convolving the extinction due
to the $\lambda 2175$ bump has created an ``artificial" kink in the
spectrum presented in this Appendix (this equates to 865 \AA\, rest
frame). A larger value of $R_{V}$ will remove this artificial bend
and these values have been found for various lines of sight in the
Galaxy \cite{car89}. Using $R_{V}$ = 4.5 or $R_{V}$ = 5.5 makes the
continuum look more like a power-law in the spectra above. Even
though the largest value of $R_{V}$ gives the smoothest continuum
fit, and improves the scatter plots in Figures 3 and 4, a
conservative intermediate choice is taken with a large uncertainty
assigned in Table 2.

0959+6827. Fortunately, this weak radio source was observed the VLA
\cite{lan06}. This is a triple radio source. Such weak sources are
very rarely imaged with high resolution VLA (as is required at
intermediate and high redshift in order to resolve most radio
sources).

1022+194. This source has a strong core (480 mJy) and a large amount
of diffuse, elongated emission (160 mJy) at 1.4 GHz in FIRST images.
The core spectrum is flat from 1.4 GHz to 5 GHz \cite{hut88}. The
total spectrum is very steep at low frequency, so it seems that the
diffuse extended emission dominates at 178 MHz. Thus, the total 178
MHz flux density is used in the estimate of Q.

1040+123. 3C 245 is the rare quasar with a 1 Jy radio core and very
strong symmetric radio lobes. In order to get a good estimate of Q,
a high resolution radio map is needed to extract the radio core and
jetted emission \cite{mur93}. Even though the radio core exceeds 1
Jy, the optical polarization and variability are small and the
emission lines have large equivalent widths, Thus, Doppler boosted
jet emission in the EUV is considered negligible.

1137+660. 3C 263 is a lobe dominated quasar, so the low frequency
total flux density at 151 MHz is the best estimator for Q.

1229-021. This is a classical triple that is steep spectrum at low
frequency. Thus, the 160 MHz flux density is used to estimate Q.
This is a low optical polarization quasar with modest optical
variability and the emission lines have large equivalent widths. It
is concluded that the continuum in the EUV is dominated by the
optically thick thermal emission. This is corroborated by negligible
changes in the EUV continuum between HST observations separated by
more than 2 years.

1241+176. A faint, very distant secondary is connected to the radio
core by a thin bridge in the 5 GHz radio image \cite{kel94}. The
flux density of this component as determined by the NVSS image at
1.4 GHz is used to estimate Q. HST observations separated by 9 years
show negligible EUV variability. This combined with large emission
line equivalent widths indicate that the radio core does not
contribute significantly to the EUV continuum.

1244+324. This is a classical triple, lobe dominated radio source.
Thus, the 151 MHz data is used to estimate Q.

1252+119. This is a core dominated radio source with a weak jet.
However, the 1.4 GHz radio image shows diffuse emission beyond the
end of the jet that is considered evidence of a core-halo type
source \cite{pri93}. The faint extension is used to estimate the
halo (lobe) flux density for use in the computation of Q. This is a
low optical polarization source that is not highly variable in the
optical. There is no EUV variability over a 4 year span of HST
observations. Thus, it is concluded that even though the source is
core dominated, the relativistic jet contributes negligibly to the
EUV.

1340+289. At 1.4 GHz, the flux is equally split between a ``core"
and the southern lobe \cite{hut88}. The 4.8 GHz image shows a small
northern extension, likely part of a lobe emission. The overall size
is at least 25 kpc at 4.8 GHz making it suitable for this sample and
the estimation techniques for Q \cite{pri93}. The spectrum is steep
at low frequency, so it is assumed that the lobe emission dominates
the 151 MHz flux density and this is used to estimate Q.  The
optical polarization is low and there is no evidence of strong
optical variability. Combined with the large equivalent widths of
the emission lines implies that there is very little synchrotron
contamination of the EUV continuum.

1340+606. 3C 288.1 is a lobe dominated radio source, so the 151 MHz
flux density is used to estimate Q.

1354+19. This source has a strong radio core and a strong one-sided
jet, similar to PKS 0637 -752. Since the counter-lobe is more
luminous than the lobe on the jetted side in the 1.4 GHz high
resolution radio images, it is safe to assume that Doppler  beaming
is insignificant in the jetted lobe \cite{mur93}. Thus, the total
lobe flux density at 1.4 GHz (less the Doppler beamed one - sided
kpc jet emission) is used to compute Q. This object has low optical
polarization and is mildly variable in the optical. Comparing the
G160L and G270H spectra taken 14 months apart, there is no
difference in the overlap region, thus the rest frame far UV flux is
not highly variable. Thus, it is concluded, in spite of the strong
radio core, that the EUV is predominantly optically thick thermal
emission.

 1415+172. This is a lobe dominated triple \cite{hut88}. Thus, the
total low frequency flux density is used to estimate Q.

1857+566. This is a powerful steep spectrum triple \cite{bar88}.
Thus, the low frequency total flux density is used for estimating Q.

2149+212. This is a compact ($20$ kpc) steep spectrum triple
\cite{bar88}. Thus, the low frequency total flux density is used for
estimating Q.

2340-036. This source is a symmetric core dominated triple radio
source in FIRST images. The 1.4 GHz flux density of the lobes is
extracted from the FIRST data and is used in the computation of Q.
Even though the radio source is core-dominated., the optical
polarization and variability are small and the emission lines have
high equivalent width, Thus, Doppler boosted jet emission in the EUV
is considered negligible.

\section*{Appendix 2.: HST Spectra}
Figure 5 are the EUV spectra corrected for Galactic extinction, the
Lyman $\alpha$ forest and Lyman limit systems as discussed in the
Sample Selection section. The spectra are arranged in order of
increasing $Q/L_{\mathrm{bol}}$ in order to see the trend of
increasing EUV deficit regardless of the precise continuum power law
fits (the black lines). The spectra are log-log plots of
$F_{\lambda}$ as a function of $\lambda$. The data was downloaded
from MAST and smoothed as required to enhance the definition of the
continuum
\begin{figure*}
\includegraphics[width=125 mm, angle= 0]{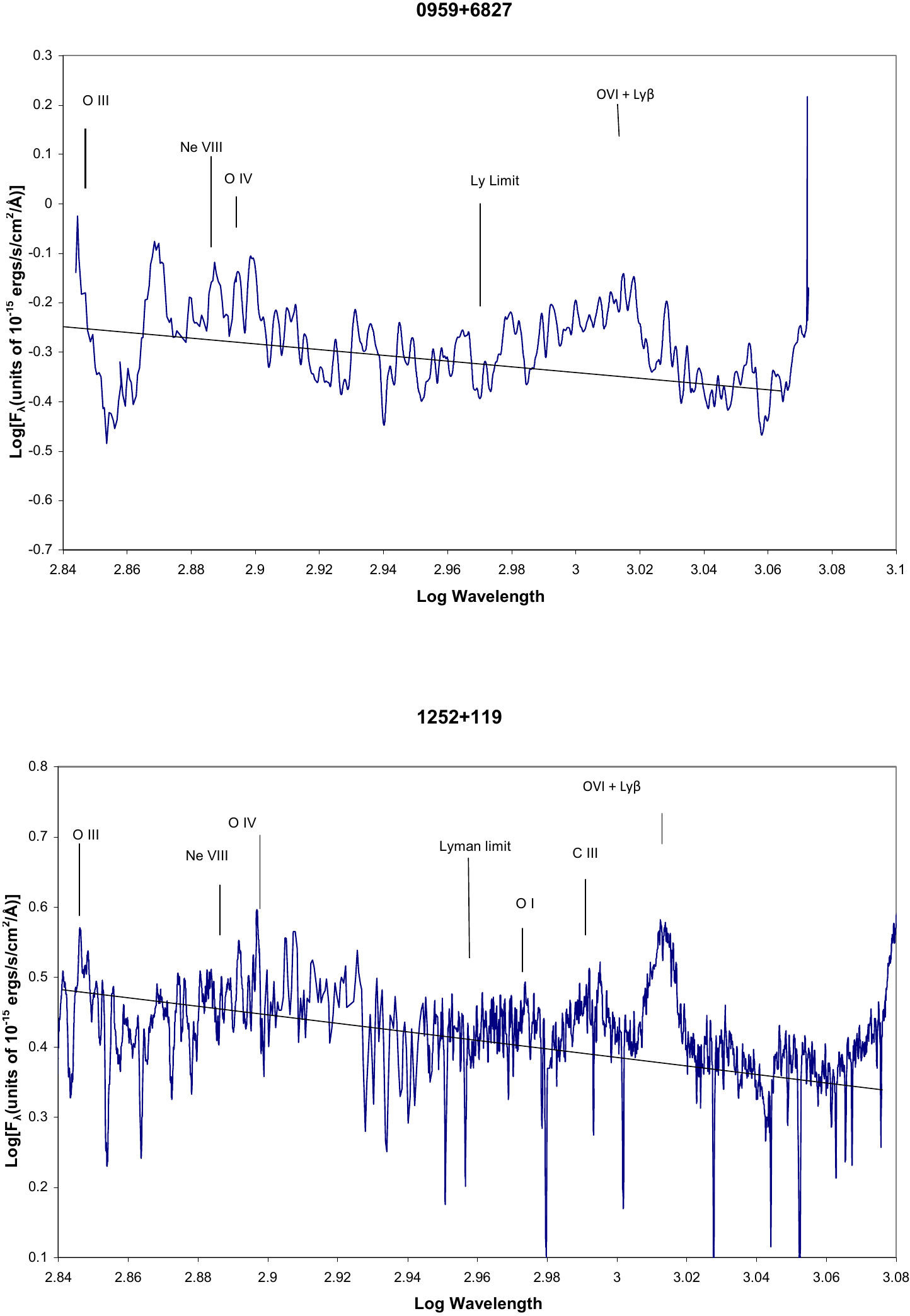}
\caption{a}
\end{figure*}
\setcounter{figure}{4}
\begin{figure*}
\includegraphics[width=125 mm, angle= 0]{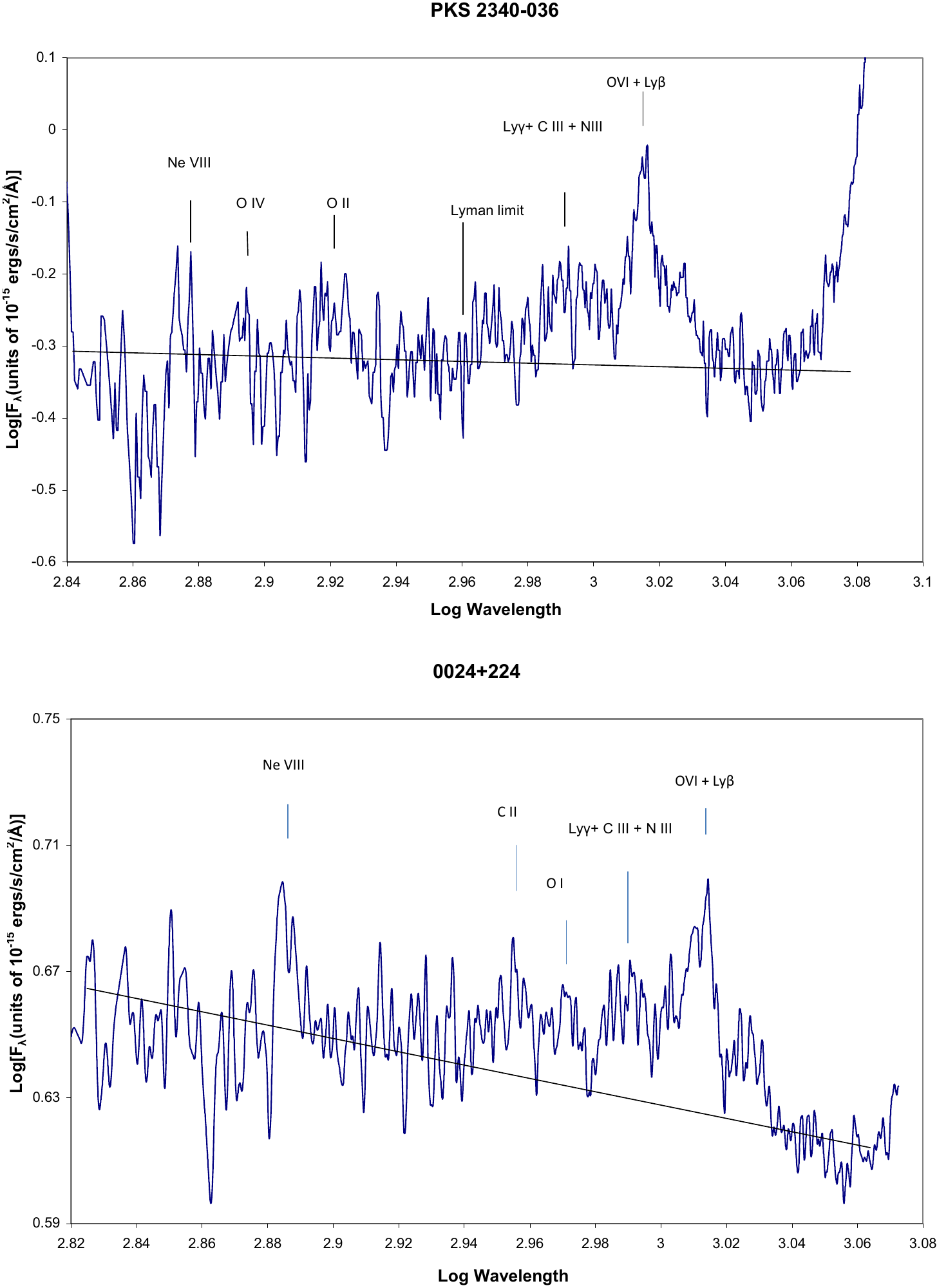}
\caption{b}
\end{figure*}
\setcounter{figure}{4}
\begin{figure*}
\includegraphics[width=125 mm, angle= 0]{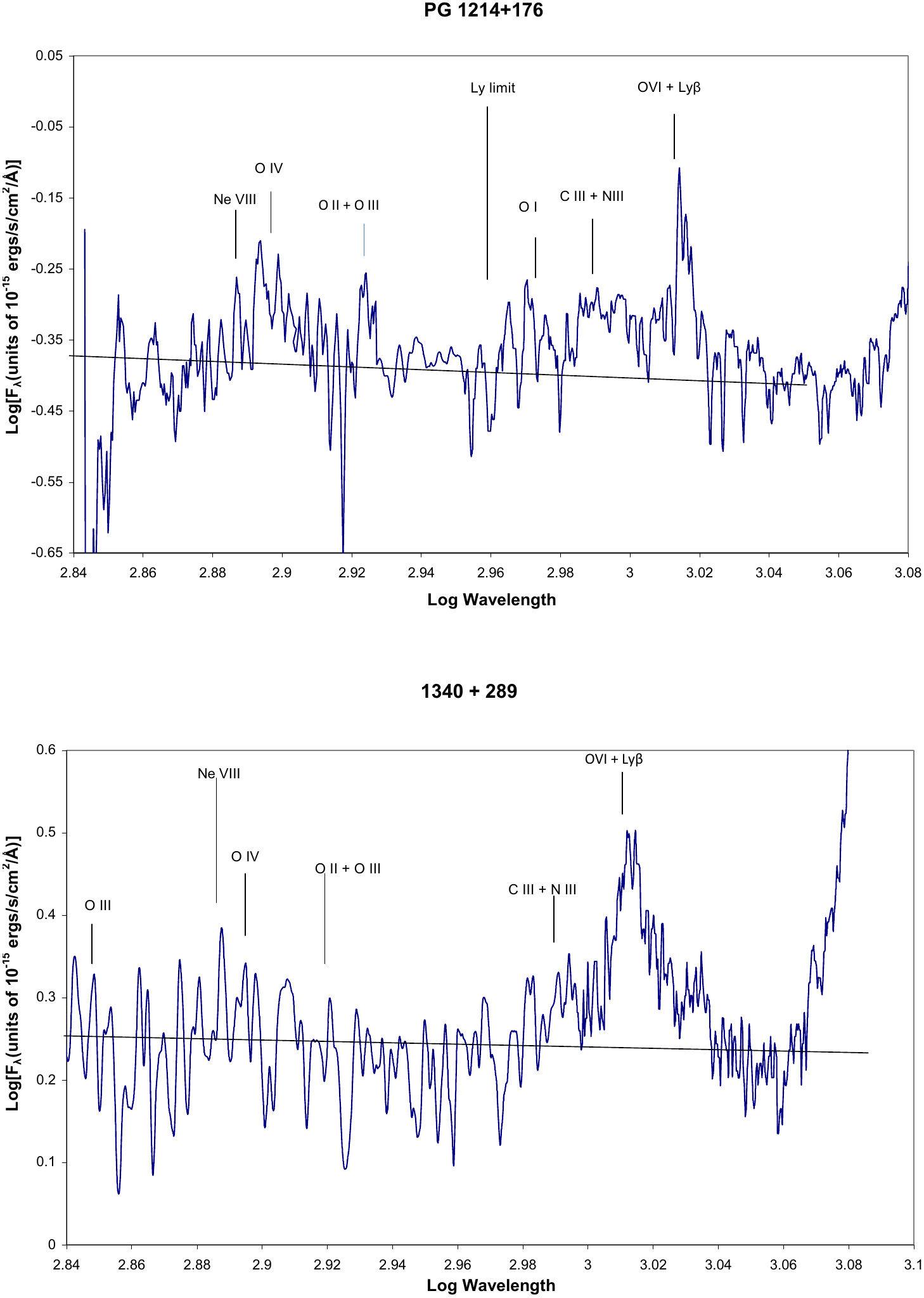}
\caption{c}
\end{figure*}
\setcounter{figure}{4}
\begin{figure*}
\includegraphics[width=125 mm, angle= 0]{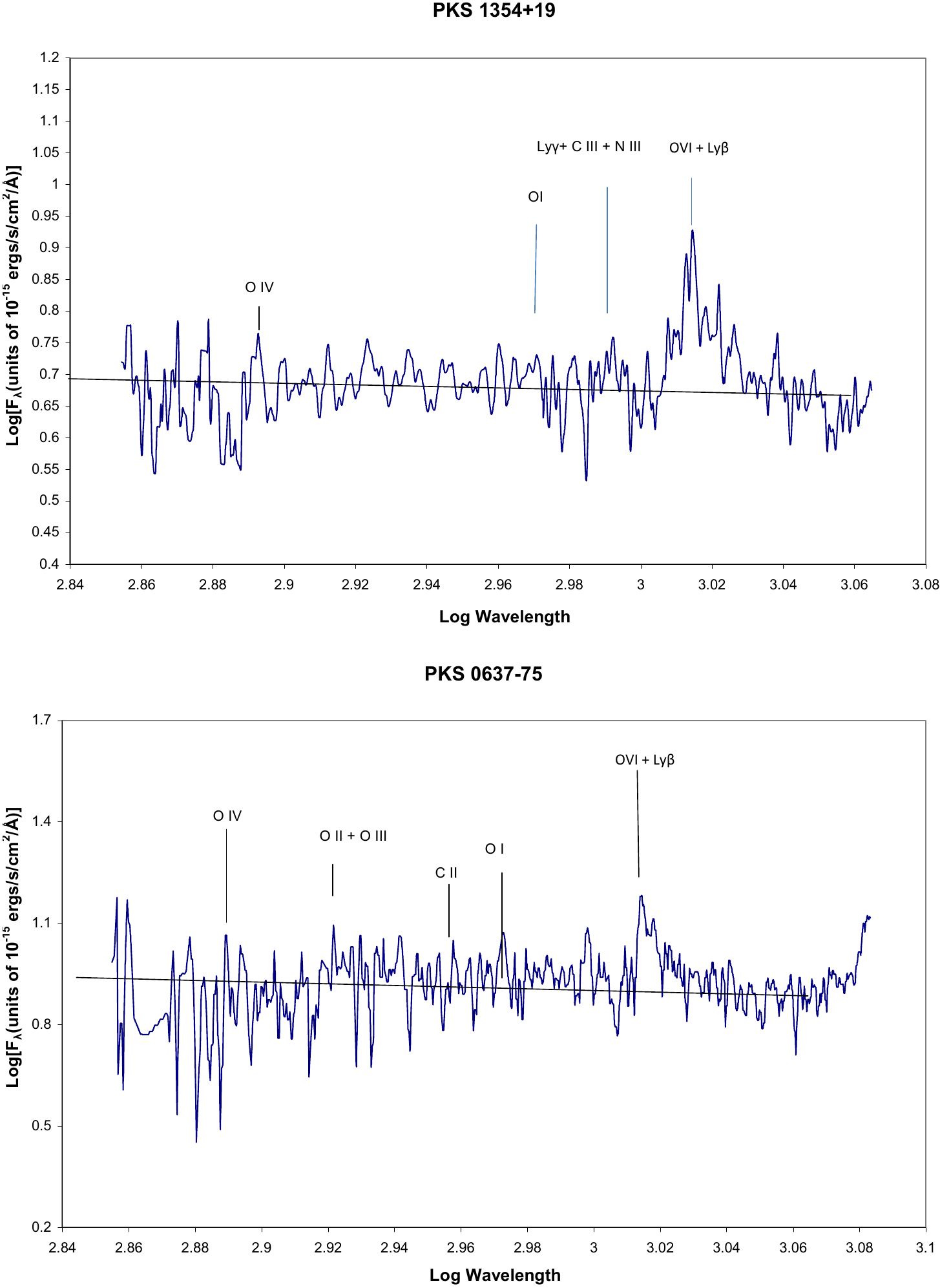}
\caption{d}
\end{figure*}
\setcounter{figure}{4}
\begin{figure*}
\includegraphics[width=125 mm, angle= 0]{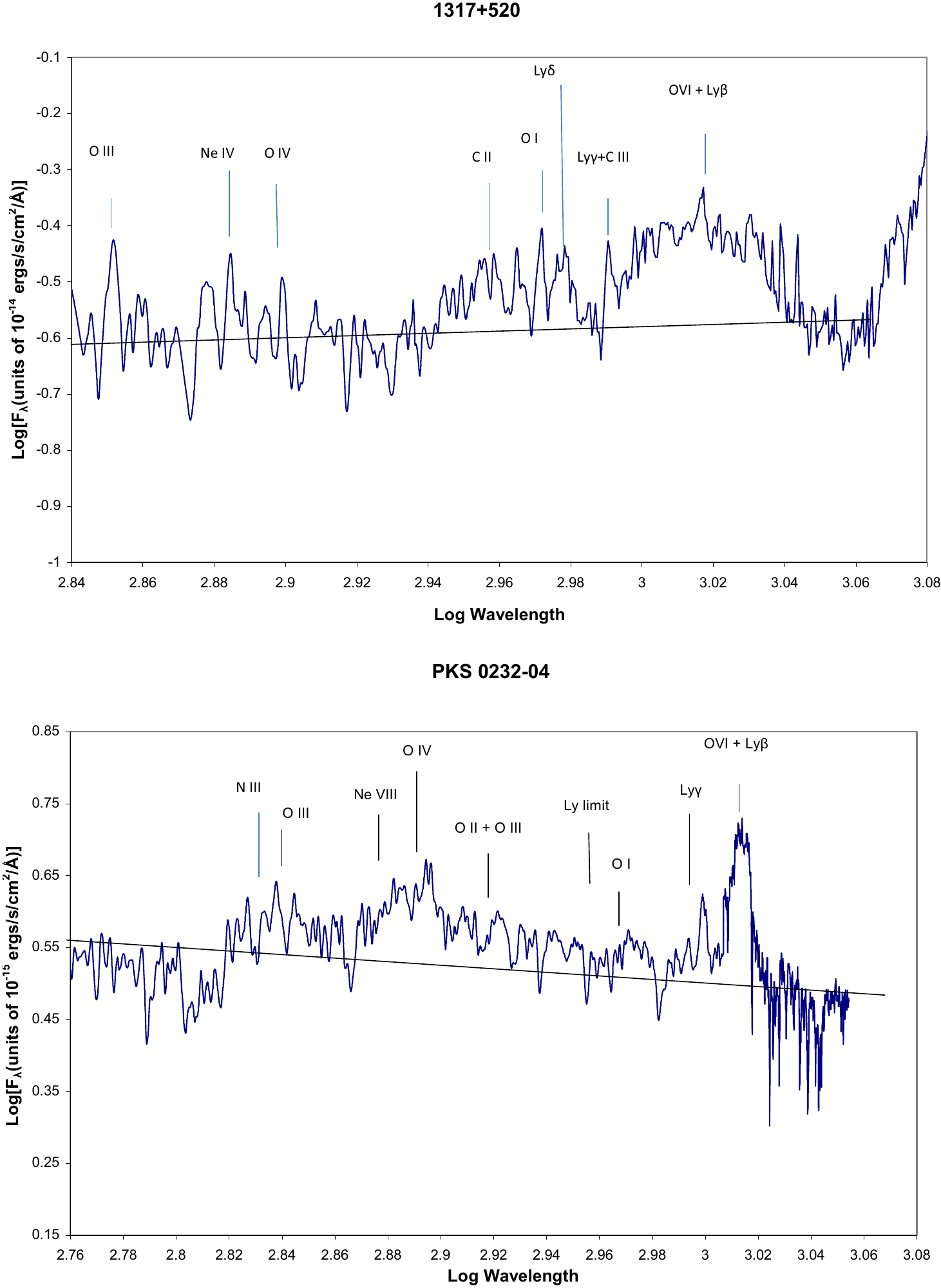}
\caption{e}
\end{figure*}
\setcounter{figure}{4}
\begin{figure*}
\includegraphics[width=125 mm, angle= 0]{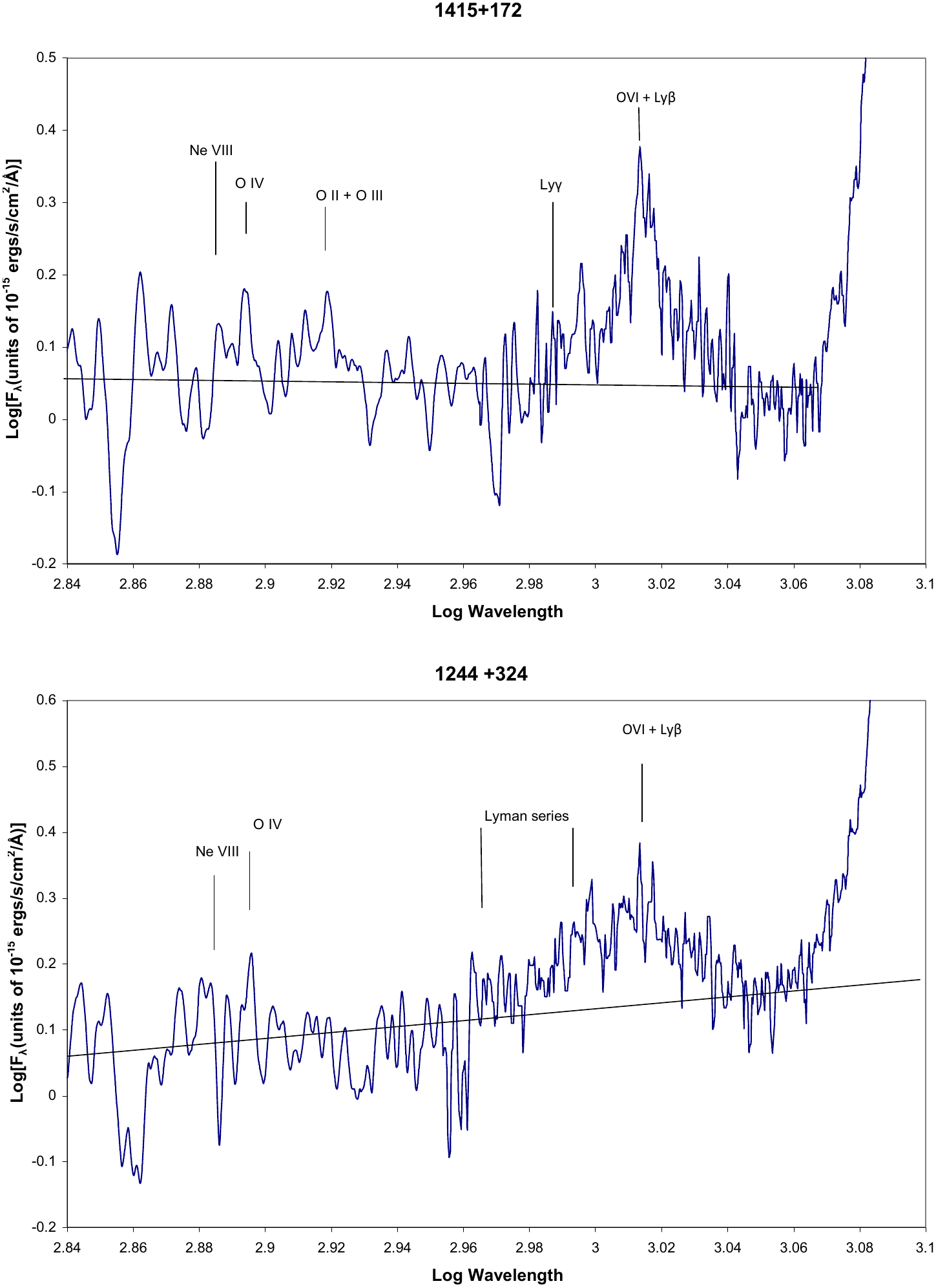}
\caption{f}
\end{figure*}
\setcounter{figure}{4}
\begin{figure*}
\includegraphics[width=125 mm, angle= 0]{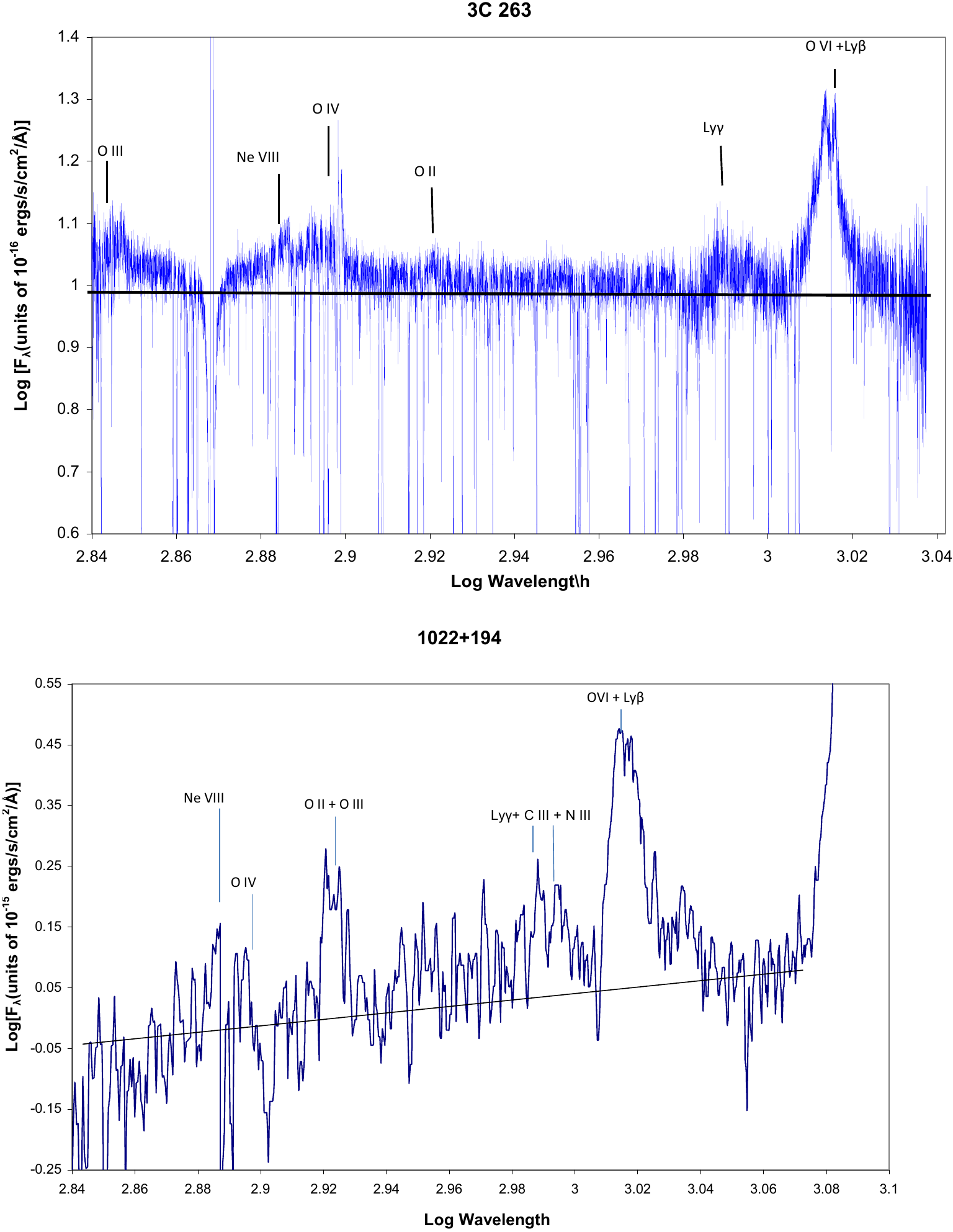}
\caption{g. The spectral data for 3C 263 was generously provided by
Michael Shull and Matt Stevans}
\end{figure*}
\setcounter{figure}{4}
\begin{figure*}
\includegraphics[width=125 mm, angle= 0]{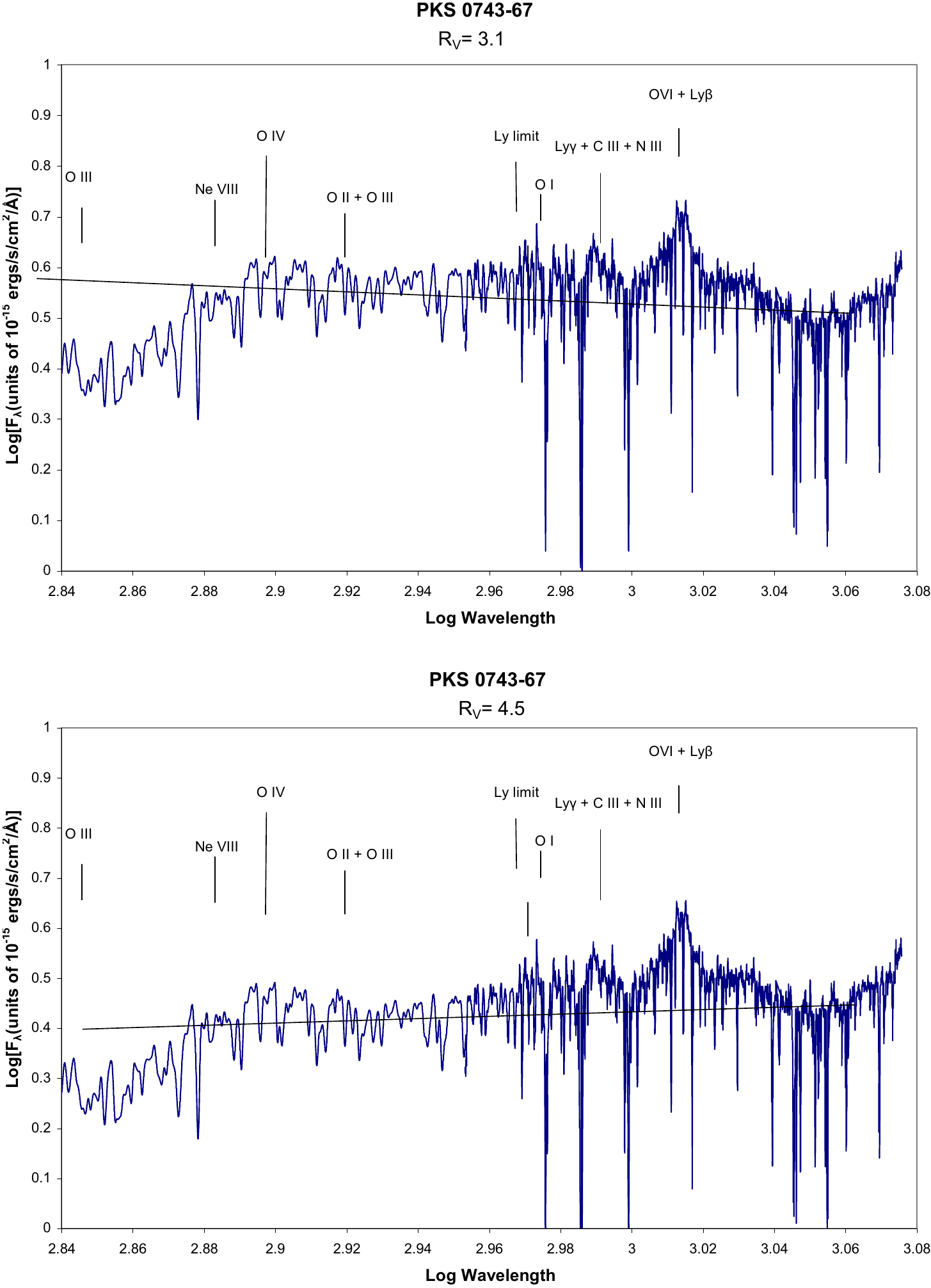}
\caption{h.}
\end{figure*}
\setcounter{figure}{4}
\begin{figure*}
\includegraphics[width=125 mm, angle= 0]{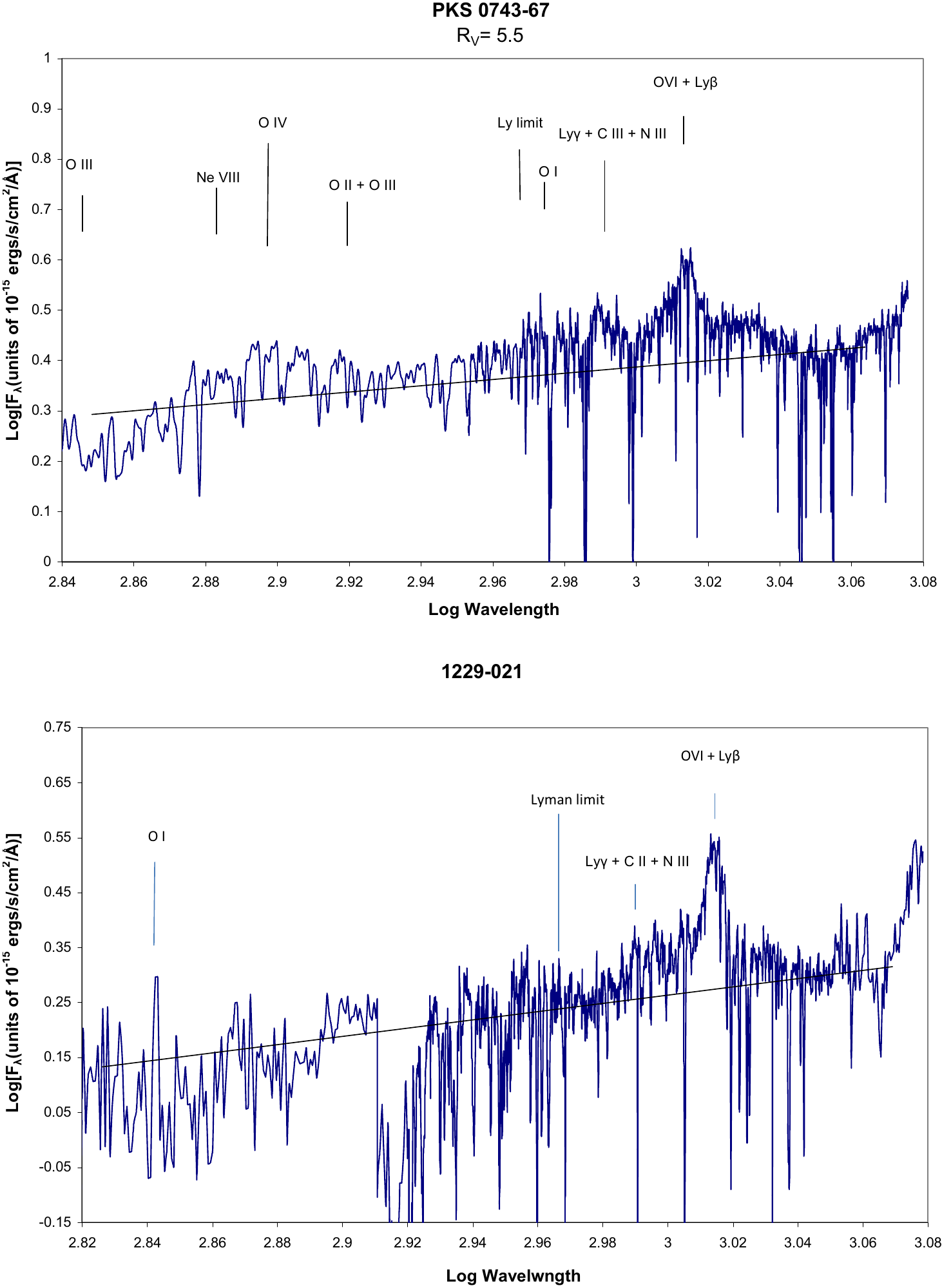}
\caption{i.}
\end{figure*}
\setcounter{figure}{4}
\begin{figure*}
\includegraphics[width=125 mm, angle= 0]{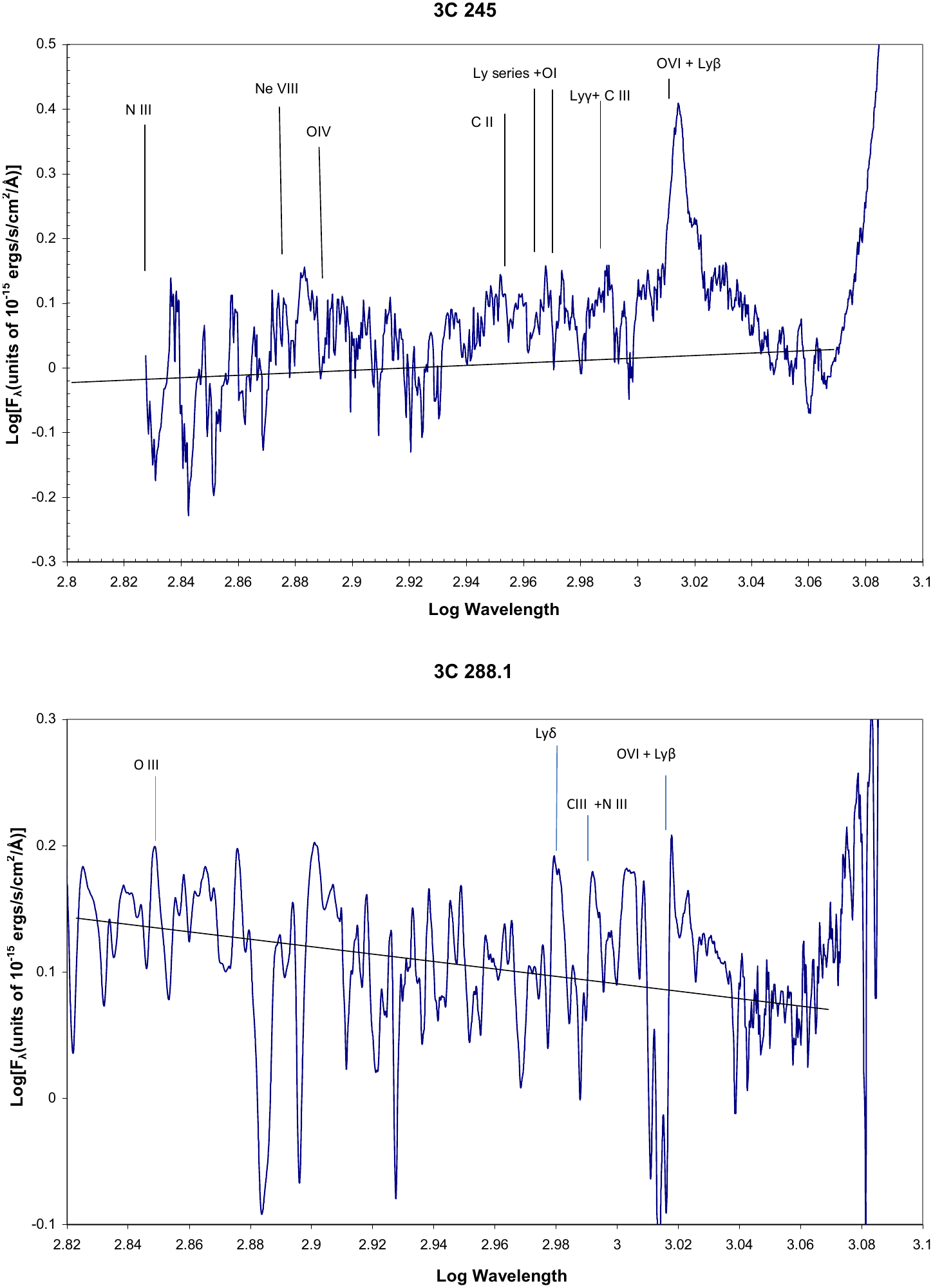}
\caption{j.}
\end{figure*}
\setcounter{figure}{4}
\begin{figure*}
\includegraphics[width=125 mm, angle= 0]{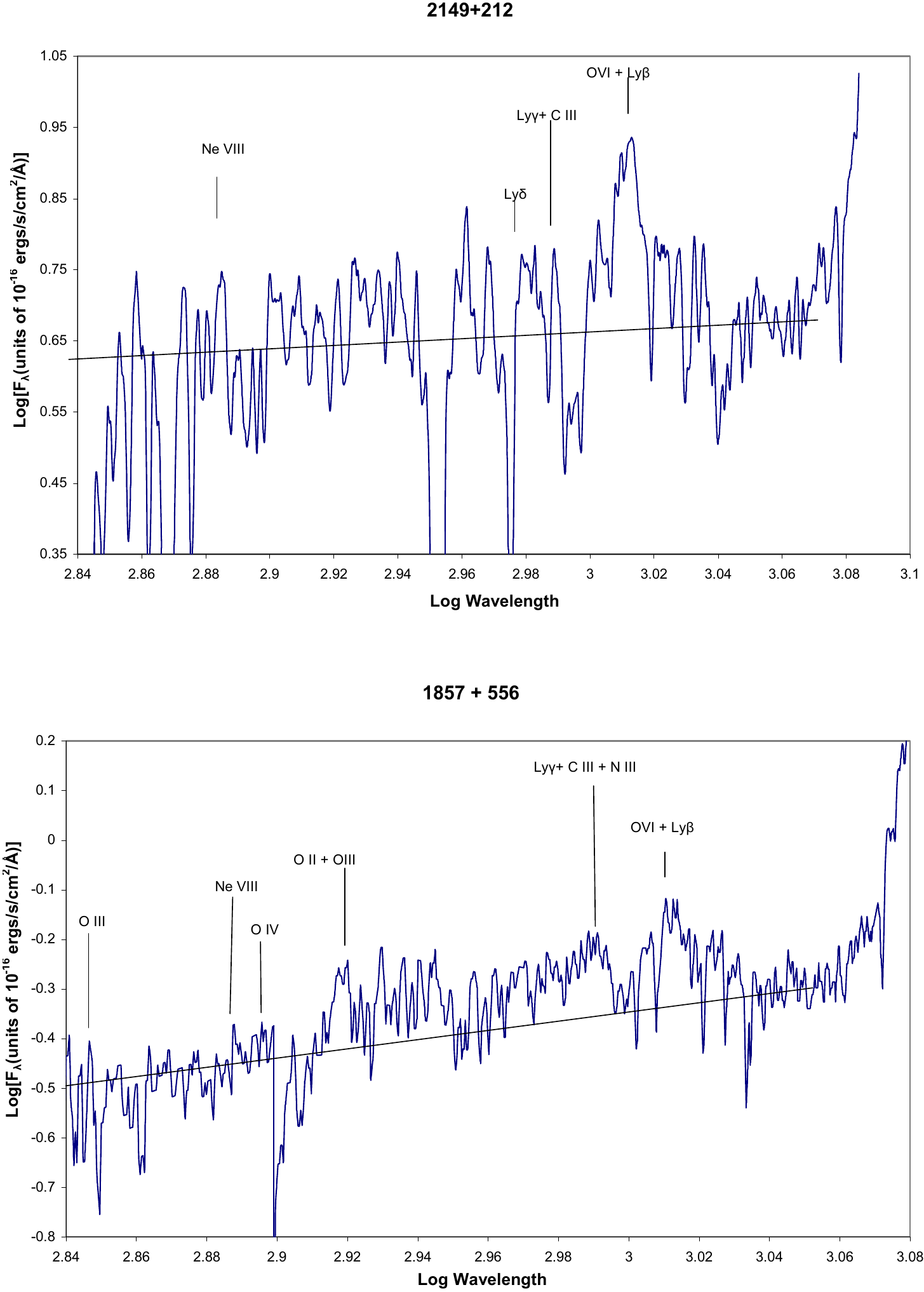}
\caption{k.}
\end{figure*}
\end{document}